\documentclass[11pt,a4paper]{article}
\pdfoutput=1 

\usepackage{jheppub}
\usepackage{latexsym}
\usepackage{revsymb}
\usepackage{bm}
\usepackage{subfigure}

\newcommand{\bra}[1]{\left\langle #1 \right|}
\newcommand{\ket}[1]{\left| #1 \right\rangle}
\newcommand{\braket}[2]{\langle #1 | #2 \rangle}
\renewcommand{\tilde}[1]{\overset{\lower1pt\hbox{$\scriptstyle{\sim}$}}{#1}}
\newcommand{\edlit}[1]{\overset{\lower1pt\hbox{$\scriptstyle{\backsim}$}}{#1}}

\preprint{IPMU13-0030}

\title{Analytical Approximation of the Neutrino Oscillation Matter Effects at large $\bm{\theta_{13}}$}

\author[a,1]{Sanjib Kumar Agarwalla,\note{Corresponding author.}}
\author[b]{Yee Kao,}
\author[c,d]{and Tatsu Takeuchi$\,$}

\affiliation[a]{Institute of Physics, Sachivalaya Marg, Sainik School Post, Bhubaneswar 751005, Orissa, India}
\affiliation[b]{Department of Chemistry and Physics, Western Carolina University, Cullowhee, NC 28723, USA}  
\affiliation[c]{Center for Neutrino Physics, Physics Department, Virginia Tech, Blacksburg, VA 24061, USA}
\affiliation[d]{Kavli Institute for the Physics and Mathematics of the Universe (WPI),
The University of Tokyo, Kashiwa-shi, Chiba-ken 277-8583, Japan}

\emailAdd{sanjib@iopb.res.in}
\emailAdd{ykao@email.wcu.edu}
\emailAdd{takeuchi@vt.edu}

\abstract{ 
We argue that the neutrino oscillation probabilities in matter
are best understood by allowing the mixing angles and 
mass-squared differences in the standard parametrization
to `run' with the matter effect parameter
$a=2\sqrt{2}G_F N_e E$, where $N_e$ is the electron density in matter and
$E$ is the neutrino energy.
We present simple analytical approximations to 
these `running' parameters. We show that for 
the moderately large value of $\theta_{13}$, as discovered by the reactor experiments,
the running of the mixing angle $\theta_{23}$ and the CP violating phase $\delta$
can be neglected. It simplifies the analysis of the resulting
expressions for the oscillation probabilities considerably.
Approaches which attempt to directly provide approximate analytical expressions for 
the oscillation probabilities in matter suffer in accuracy
due to their reliance on expansion in $\theta_{13}$, 
or in simplicity when higher order terms in $\theta_{13}$ are included.
We demonstrate the accuracy of our method by comparing it to the exact numerical result, 
as well as the direct approximations of Cervera et al., Akhmedov et al., Asano and Minakata, 
and Freund.
We also discuss the utility of our approach in figuring out the
required baseline lengths and neutrino energies for the oscillation probabilities to exhibit
certain desirable features.
}

\keywords{Neutrino, Oscillation Probability, Matter Effect, Jacobi Method}
\arxivnumber{1302.6773}

\begin{document}
\maketitle
\flushbottom

\section{Introduction}
\label{sec:introduction}

When performing long-baseline neutrino oscillation experiments on the Earth
with accelerator based beams,
or when detecting atmospheric neutrinos coming from below, 
the neutrinos necessarily traverse the Earth's interior \cite{Minakata:1998bf,Banuls:2001zn,Huber:2002mx,Gandhi:2004bj,Huber:2005ep,Akhmedov:2006hb,Agarwalla:2012uj,Blennow:2013rca}.
This makes the
understanding of matter effects \cite{Wolfenstein:1977ue,Mikheev:1986gs,Mikheev:1986wj,Barger:1980tf} 
on the neutrino oscillation probabilities
an indispensable part of analyzing such experiments.
These matter effects can of course be calculated numerically for arbitrary matter profiles,
but approximate analytical expressions are useful not only for making initial estimates
on the requirements one must place on long-baseline experiments, but 
in obtaining a deeper understanding of the physics involved.

The exact three-flavor neutrino oscillation probabilities in constant-density matter 
can be expressed analytically
\cite{Barger:1980tf,Toshev:1986fs,Petcov:1986qg,Kim:1986vg,Zaglauer:1988gz,Krastev:1988yu,Toshev:1991ku,Dick:1999ed,Ohlsson:1999xb,Ohlsson:1999um,Freund:2001pn,Kimura:2002wd}.
This requires the diagonalization of the $3\times 3$ 
effective Hamiltonian in matter whose $ee$-element in the
flavor basis is shifted by $a=2\sqrt{2}G_F N_e E$,
where $N_e$ is the electron density and $E$ is the neutrino energy.
The eigenvalues of the effective Hamiltonian yield the effective neutrino mass-squared differences in matter\footnote{The cubic characteristic equation for the eigenvalues of the effective Hamiltonian can be solved analytically using Cardano's formula \cite{Cardano:1545}.},
while the diagonalization matrix is multiplied with the
vacuum neutrino mixing matrix to yield its in-matter counterpart.
Many authors adopt the standard vacuum parameterization of the mixing matrix to the matter version,
and absorb matter effects into shifts of 
the mixing angles and CP violating phase, yielding the effective values of these parameters in matter
\cite{Toshev:1986fs,Petcov:1986qg,Zaglauer:1988gz,Freund:2001pn}.
Thus, the neutrino oscillation probabilities in matter can be obtained from those in vacuum by 
simply replacing the mass-squared differences, mixing angles, and CP violating phase with their
effective values.
Unfortunately, the final exact expressions for the
neutrino oscillation probabilities obtained this way are too complicated to yield physical insight,
especially if re-expressed in terms of the vacuum parameters.

Consequently, various analytical approximations have been devised to
better understand the physics potential of various neutrino experiments \cite{Barger:1980tf,Kuo:1986sk,Arafune:1997hd,Cervera:2000kp,Freund:2001pn,Peres:2003wd,Akhmedov:2004ny,Akhmedov:2004rq,Akhmedov:2008nq,Asano:2011nj}.
These approximations relied on expansions in the small parameters 
$\alpha=\delta m^2_{21}/\delta m^2_{31}\approx 0.03$ and/or $s_{13}=\sin\theta_{13}$
in one form or another, a systematic treatment of which can be found in Ref.~\cite{Akhmedov:2004ny}.
In some cases the matter-effect parameter $a=2\sqrt{2}G_F N_e E$
was also assumed to be small \cite{Barger:1980tf,Arafune:1997hd}.
For instance, the formula of
Cervera et al. in Ref.~\cite{Cervera:2000kp} 
and that of Ahkmedov et al. in Ref.~\cite{Akhmedov:2004ny}
include terms of order $O(\alpha^2)$, $O(\alpha s_{13})$, and $O(s_{13}^2)$.
Unfortunately, the accuracies of these formulae suffer when the value of
$\theta_{13}$ is as large as was measured by 
Daya Bay \cite{An:2012eh,An:2012bu} and RENO \cite{Ahn:2012nd},
consistent with both earlier and later determinations by
T2K \cite{Abe:2011sj}, MINOS \cite{Adamson:2011qu,Adamson:2013ue}, and
Double Chooz \cite{Abe:2011fz,Abe:2012tg}.
Given that the current world average of $s_{13}=\sin\theta_{13}$ is about $0.15$ \cite{GonzalezGarcia:2012sz}, the terms included are not all of the same order.
Asano and Minakata \cite{Asano:2011nj} have derived 
the order $O(\alpha s_{13}^2)$ and $O(s_{13}^4)$ corrections to the Cervera et al. formula,
but the simplicity of the original expressions is lost.
Further improvements in accuracy are possible at the expense of simplicity, as was shown by 
Freund in Ref.~\cite{Freund:2001pn} where an approximate expression for the oscillation probability $P(\nu_e\rightarrow\nu_\mu)$ including all orders of $\theta_{13}$ was derived.

In previous papers \cite{Honda:2006hp,Honda:2006gv},
we had argued the advantage of not expressing the neutrino oscillation probabilities in 
matter directly in terms of the vacuum parameters, 
but to maintain their expressions in terms of the effective parameters
in matter which `ran' with the parameter $a=2\sqrt{2}G_F N_e E$.
Further, it was shown that the Jacobi method \cite{Jacobi:1846} could be utilized 
to find approximate expressions for the `running' parameters in a systematic fashion,
leading to fairly simple and compact expressions.
In particular, it was shown that the effective values of $\theta_{23}$ and the CP violating phase $\delta$
do not `run' to the order considered, maintaining their vacuum values
at all neutrino energies and baselines.
(The non-running of $\theta_{23}$ and $\delta$ has also been
discussed in Ref.~\cite{Krastev:1988yu}.)
The $a$-dependence of the resulting expressions for the oscillation probabilities
in terms of the approximate running parameters could be analyzed in a simple manner, 
facilitating the understanding of matter effects.

The approximation of Refs~\cite{Honda:2006hp,Honda:2006gv} worked extremely well 
except when $\theta_{13}$ was very small, a possibility that could not be ignored until the Daya Bay/RENO measurements.
In this paper, we reintroduce the method with further refinements which
improve the accuracy of the approximation for large $\theta_{13}$, while maintaining its ease of use.

This paper is organized as follows. 
In section 2, we explain our approach to the matter effect problem, and list all
the formulae necessary to calculate the approximate `running' parameters in our approach.
Approximate oscillation probabilities are obtained by replacing the 
mass-squared differences and mixing angles in the vacuum oscillation probabilities
with their effective `running' values.
In section 3, we demonstrate the accuracy of our approximation 
at various baseline lengths, different mass hierarchies, and different values of the CP violating phase $\delta$.
Comparisons with the approximations of Cervera et al. \cite{Cervera:2000kp},
Akhmedov et al \cite{Akhmedov:2004ny},
Asano-Minakata \cite{Asano:2011nj}, and Freund \cite{Freund:2001pn} 
are also made.
In section 4, we show how simple calculations
using our approximation can be used to derive the baselines and energies at which
the oscillation probabilities exhibit desirable features.
We conclude in Section 5.
Detailed derivation of our approximation is given in appendices \ref{sec:Jacobi} and \ref{sec:Commute}.


\section{The Approximation}
\label{sec:approximation}

\begin{figure}[t]
\begin{center}
\includegraphics[width=7cm]{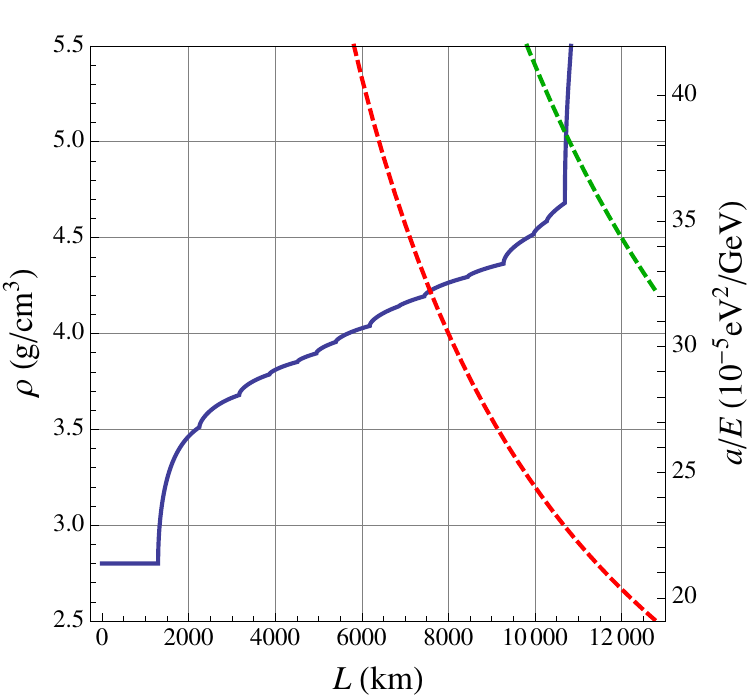}
\end{center}
\caption{The dependence of the line-averaged mass density $\rho$
on the baseline length $L$ based on the 
Preliminary Reference Earth Model \cite{PREM:1981}.
The labels on the right edge of the frame indicate
the corresponding values of $a/E$.
The green and red dashed lines indicate
$\rho L=54000\,\mathrm{km\cdot g/cm^3}$ and
$\rho L=32000\,\mathrm{km\cdot g/cm^3}$, respectively, which 
are conditions that will be
discussed in section~\ref{sec:applications}.
}
\label{rhofig}
\end{figure}

In the following, we use the conventions and notation reviewed in Appendix~\ref{sec:basics}.

\subsection{Diagonalization of the Effective Hamiltonian}
\label{subsec:effective_hamiltonian}

If the matter density along the baseline is constant\footnote{At baseline length $L=10690$ km or longer, the neutrino beam crosses the core-mantle-boundary and experiences a sudden jump in matter density. See Ref. \cite{Liao:2007re} for treatments of  non-adiabatic transitions.}, 
the effective Hamiltonian which governs the evolution
of neutrino flavor in matter is given by
\begin{equation}
H_a \;=\; 
U
\left[ \begin{array}{ccc} 0 & 0 & 0 \\
                          0 & \delta m^2_{21} & 0 \\
                          0 & 0 & \delta m^2_{31}
       \end{array}
\right]
U^\dagger +
\left[ \begin{array}{ccc} a & 0 & 0 \\
                          0 & 0 & 0 \\
                          0 & 0 & 0 
       \end{array}
\right] \;,
\label{Hnu}
\end{equation}
where $U$ is the neutrino mixing matrix in vacuum, and
\begin{equation}
a 
\;=\; 2\sqrt{2}\,G_F N_e E
\;=\; 7.63\times 10^{-5}\left(\mathrm{eV}^2\right)
\left(\dfrac{\rho}{\mathrm{g/cm^3}}\right)
\left(\dfrac{E}{\mathrm{GeV}}\right)\;.
\label{adef}
\end{equation}
Here, $N_e$ is the electron number density, 
$\rho$ the matter mass density along the baseline,
and $E$ the neutrino energy.
The above term appearing in the $ee$-component of $H_a$ 
is due to the interaction of $\nu_e$
with the electrons in matter via $W$-exchange,
and Eq.~(\ref{adef}) assumes $N_e=N_p=N_n$ in Earth matter.
It also assumes $E\ll M_W$ since the $W$-exchange interaction is
approximated by a point-like four-fermion interaction.
$Z$-exchange effects are flavor universal and 
only contribute a term proportional to the unit matrix
to $H_a$, which can be dropped.

If we write the eigenvalues of $H_a$ as $\lambda_i$ ($i=1,2,3$) and the diagonalization matrix as $\tilde{U}$, that is
\begin{equation}
H_a \;=\; 
\tilde{U}
\left[ \begin{array}{ccc} \lambda_1 & 0 & 0 \\
                          0 & \lambda_2 & 0 \\
                          0 & 0 & \lambda_3
       \end{array}
\right]
\tilde{U}^\dagger
\;,
\label{Utildelambdadef}
\end{equation}
then the neutrino oscillation probabilities in matter
are obtained by simply taking their expressions in vacuum
and replacing the elements of the mixing matrix $U$ and the
mass-square differences $\delta m^2_{ij}$ 
with their effective `running' values in matter
\cite{Wolfenstein:1977ue,Mikheev:1986gs,Mikheev:1986wj}
:
\begin{equation}
U_{\alpha i}\;\rightarrow\; \tilde{U}_{\alpha i}\;,
\qquad
\delta m^2_{ij}\;\rightarrow\; 
\delta \lambda_{ij}\;\equiv\;\lambda_i-\lambda_j\;.
\end{equation}
Note that $a$
is $E$-dependent, which means that both $\tilde{U}_{\alpha i}$ and
$\delta\lambda_{ij}$ are also $E$-dependent.
They also depend on the baseline length $L$ 
since the average matter density $\rho$ along a baseline 
varies with $L$.
The $L$-dependence of the average $\rho$ and 
the corresponding value of $a/E$ are 
shown in Fig.~\ref{rhofig}.

For anti-neutrino beams, 
the flavor-evolution Hamiltonian in matter is
\begin{equation}
\overline{H}_a 
\;=\; U^*
\left[ \begin{array}{ccc} 0 & 0 & 0 \\
                          0 & \delta m^2_{21} & 0 \\
                          0 & 0 & \delta m^2_{31}
       \end{array}
\right]
U^\mathrm{T} +
\left[ \begin{array}{ccc} -a & 0 & 0 \\
                          0 & 0 & 0 \\
                          0 & 0 & 0 
       \end{array}
\right] \;.
\label{Hnubar}
\end{equation}
In comparison to Eq.~(\ref{Hnu}), 
the CP violating phase $\delta$ in $U$ and the
matter-effect term $a$ both acquire minus signs.
Let us write the eigenvalues of $\overline{H}_a$ as $\overline{\lambda}_i$ ($i=1,2,3$)
and the diagonalization matrix as $\edlit{U}$,
that is
\begin{equation}
\overline{H}_a 
\;=\; 
\edlit{U}^*
\left[ \begin{array}{ccc} \overline{\lambda}_1 & 0 & 0 \\
                          0 & \overline{\lambda}_2 & 0 \\
                          0 & 0 & \overline{\lambda}_3
       \end{array}
\right]
\edlit{U}^\mathrm{T}
\;.
\label{Uedlitlambdabardef}
\end{equation}
Note that the tilde above $\edlit{U}$ here is flipped 
to distinguish it from $\tilde{U}$ in Eq.~(\ref{Utildelambdadef}).
The anti-neutrino oscillation probabilities in matter are then
obtained by making the replacements
\begin{equation}
U_{\alpha i}\;\rightarrow \edlit{U}_{\alpha i}\;,\qquad
\delta m^2_{ij} \;\rightarrow\; \delta\overline{\lambda}_{ij}
\;\equiv\;\overline{\lambda}_i-\overline{\lambda}_j\;,
\end{equation}
in the vacuum expressions.

\subsection{Effective Running Mixing Angles}
\label{subsec:effective_mixing_angles}

While it is possible to write down exact analytical
expressions for $\tilde{U}_{\alpha i}$ and $\delta\lambda_{ij}$,
as well as their anti-neutrino counterparts \cite{Zaglauer:1988gz}, 
simpler and more transparent approximate expressions are often desirable. 
One popular approach is to expand the probability formulae in terms of small 
parameters such as $\delta m^2_{21}/|\delta m^2_{31}|$ and $\theta_{13}$.
Our approach, however, utilizes the Jacobi method \cite{Jacobi:1846}. Instead of
obtaining approximations for the probabilities directly,
we derived the approximations for the effective mixing parameters.
In the following two sections, we list the expressions necessary to calculate the effective running mixing angles and the effective running mass-squared differences for the neutrino and anti-neutrino cases separately.
Detailed derivation of our approximation is given in Appendix~\ref{sec:Jacobi}.

\subsection{Neutrino Case}

We first recognize that the mixing matrix in matter can be parameterized in the
same fashion as in the vacuum case:
\begin{eqnarray}
\tilde{U} \;=\;
R_{23}(\tilde{\theta}_{23},0)
R_{13}(\tilde{\theta}_{13},\tilde{\delta})
R_{12}(\tilde{\theta}_{12},0)\;.
\end{eqnarray}
The effective mixing angles can be approximated by
\begin{eqnarray}
\tilde\theta_{12} & \approx & \theta'_{12} \;,\cr
\tilde\theta_{13} & \approx & \theta'_{13} \;,\cr
\tilde\theta_{23} & \approx & \theta_{23} \;,\cr
\tilde\delta & \approx & \delta \;,
\end{eqnarray}
where $\theta'_{12}$ and $\theta'_{13}$ are given by
\begin{eqnarray}
\tan 2\theta'_{12} 
& = & 
\dfrac{(\delta m^2_{21}/c_{13}^2)\sin 2\theta_{12}}
      {(\delta m^2_{21}/c_{13}^2)\cos 2\theta_{12} - a}
\;, 
\cr
\tan 2\theta'_{13} 
& = &
\dfrac{(\delta m^2_{31}-\delta m^2_{21}s_{12}^2)\,\sin 2\theta_{13}}
      {(\delta m^2_{31}-\delta m^2_{21}s_{12}^2)\,\cos 2\theta_{13}-a}
\;.
\label{effectivethetadef}
\end{eqnarray}
while the angle $\theta_{23}$ and the CP-violating phase $\delta$ at kept at their vacuum values \cite{Krastev:1988yu}.

The eigenvalues $\lambda_i$ ($i=1,2,3$) of $H_a$ are also given approximate running expressions:
\begin{eqnarray}
\lambda_1 & \approx & \lambda'_{-} \;,\cr
\lambda_2 & \approx & \lambda''_{\mp} \;,\cr
\lambda_3 & \approx & \lambda''_{\pm} \;,
\label{neutrinolambda}
\end{eqnarray}
where the upper(lower) sign is for the normal(inverted) hierarchy, with
\begin{eqnarray}
\lambda'_{\pm}
& \equiv & 
\dfrac{ (\delta m^2_{21}+a c_{13}^2)
        \pm\sqrt{ (\delta m^2_{21}-a c_{13}^2)^2 + 4 a c_{13}^2 s_{12}^2 \delta m^2_{21} }
      }
      { 2 }
\;,
\cr      
\lambda''_{\pm} 
& \equiv &
\dfrac{ \bigl[ \lambda'_{+} + (\delta m^2_{31}+a s_{13}^2) \bigr]
\pm \sqrt{ \bigl[ \lambda'_{+} - (\delta m^2_{31}+a s_{13}^2) \bigr]^2 
         + 4 a^2 s^{\prime 2}_{12}\,c_{13}^2\, s_{13}^2 }
      }
      { 2 } 
\;,
\label{lambdaprimesdef}
\end{eqnarray}
and $s_{12}^{\prime 2} = \sin^2\theta'_{12}$.
For the inverted hierarchy case, $\delta m^2_{31}<0$, the above expressions simplify to
\begin{equation}
\lambda_2 \;\approx\; \lambda''_+ \;\approx\; \lambda'_+\;,\qquad
\lambda_3 \;\approx\; \lambda''_- \;\approx\; \delta m^2_{31} \;<\;0\;.
\end{equation}
Thus, to take matter effects into account when calculating 
neutrino oscillation probabilities, 
all that is necessary is to take their expressions 
in terms of the mixing angles and CP-phase
in vacuum as is, and replace 
the two angles as well as the mass-squared differences with their
effective running values in matter: 
$\theta_{12}\rightarrow\theta'_{12}$, 
$\theta_{13}\rightarrow\theta'_{13}$,
$\delta m^2_{ij}\rightarrow \delta\lambda_{ij}=\lambda_i-\lambda_j$.
This simplifies the calculation considerably, and allows for a transparent understanding
of how matter-effects affect neutrino oscillation 
by looking at the $a$-dependence of the effective parameters.

\subsection{Anti-Neutrino Case}
Similarly, in the anti-neutrino case, the mixing matrix can be parameterized by:
\begin{equation}
\edlit{U} \;=\;
R_{23}(\edlit{\theta}_{23},0)
R_{13}(\edlit{\theta}_{13},\edlit{\delta})
R_{12}(\edlit{\theta}_{12},0)\;.
\end{equation}
Note that the sign in front of the matter effect parameter $a$ is flipped
relative to the neutrino case, so these effective mixing angles will be different.
Our approximation is given by
\begin{eqnarray}
{\edlit{\theta}}_{12} & \approx & \overline{\theta}'_{12} \;,\cr
{\edlit{\theta}}_{13} & \approx & \overline{\theta}'_{13} \;,\cr
{\edlit{\theta}}_{23} & \approx & \theta_{23} \;,\cr
{\edlit{\delta}} & \approx & \delta \;,
\end{eqnarray}
where
\begin{eqnarray}
\tan 2\overline{\theta}'_{12} 
& = & 
\dfrac{(\delta m^2_{21}/c_{13}^2)\sin 2\theta_{12}}
      {(\delta m^2_{21}/c_{13}^2)\cos 2\theta_{12} + a}
\;, 
\cr
\tan 2\overline{\theta}'_{13} 
& = &
\dfrac{(\delta m^2_{31}-\delta m^2_{21}s_{12}^2)\,\sin 2\theta_{13}}
      {(\delta m^2_{31}-\delta m^2_{21}s_{12}^2)\,\cos 2\theta_{13}+a}
\;.
\label{effectivethetabardef}
\end{eqnarray}
Again, $\theta_{23}$ and $\delta$ are unaffected while
$\theta_{12}$ and $\theta_{13}$ are replaced by their effective running values
in matter.

The eigenvalues $\overline{\lambda}_i$ ($i=1,2,3$) of $\overline{H}_a$ are
given approximate running expressions as in the neutrino case.
The three eigenvalues of the effective Hamiltonian are approximated by
\begin{eqnarray}
\overline{\lambda}_1 & \approx & \overline{\lambda}''_{\mp} \;,\cr
\overline{\lambda}_2 & \approx & \overline{\lambda}'_{+} \;,\cr
\overline{\lambda}_3 & \approx & \overline{\lambda}''_{\pm} \;,
\end{eqnarray}
where the upper(lower) sign is for the normal(inverted) hierarchy, with
\begin{eqnarray}
\overline{\lambda}'_{\pm}
& \equiv & 
\dfrac{ (\delta m^2_{21}-a c_{13}^2)
        \pm\sqrt{ (\delta m^2_{21}+a c_{13}^2)^2 - 4 a c_{13}^2 s_{12}^2 \delta m^2_{21} }
      }
      { 2 }
\;,
\cr      
\overline{\lambda}''_{\pm} 
& \equiv &
\dfrac{ [ \overline{\lambda}'_{-} + (\delta m^2_{31}-a s_{13}^2) ]
\pm \sqrt{ [ \overline{\lambda}'_{-} - (\delta m^2_{31}-a s_{13}^2) ]^2 
         + 4 a^2 \overline{c}^{\prime 2}_{12}\,c_{13}^2\, s_{13}^2 }
      }
      { 2 } 
\;,
\label{lambdabarprimesdef}
\end{eqnarray}
and $\overline{c}_{12}^{\prime 2} = \cos^2\overline{\theta}'_{12}$.
For the normal hierarchy case, $\delta m^2_{31}>0$, the above expressions
simplify to
\begin{equation}
\overline{\lambda}_1 
\;\approx\; \overline{\lambda}''_- 
\;\approx\; \overline{\lambda}_-
\;,
\qquad
\overline{\lambda}_3 
\;\approx\; \overline{\lambda}''_+ 
\;\approx\; \delta m^2_{31}\;.
\end{equation}
Thus, the calculation of matter effects for 
anti-neutrino beams entails the replacements
$\theta_{12}\rightarrow\overline{\theta}'_{12}$, 
$\theta_{13}\rightarrow\overline{\theta}'_{13}$,
$\delta m^2_{ij}\rightarrow \delta\overline{\lambda}_{ij}=\overline{\lambda}_i-\overline{\lambda}_j$.

\begin{figure}
\begin{center}
\subfigure[neutrino mixing angles]{\includegraphics[width=7.5cm]{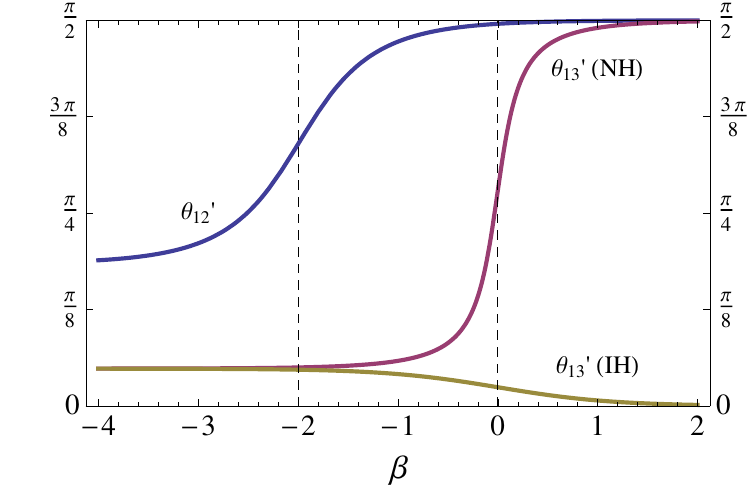}}
\subfigure[anti-neutrino mixing angles]{\includegraphics[width=7.5cm]{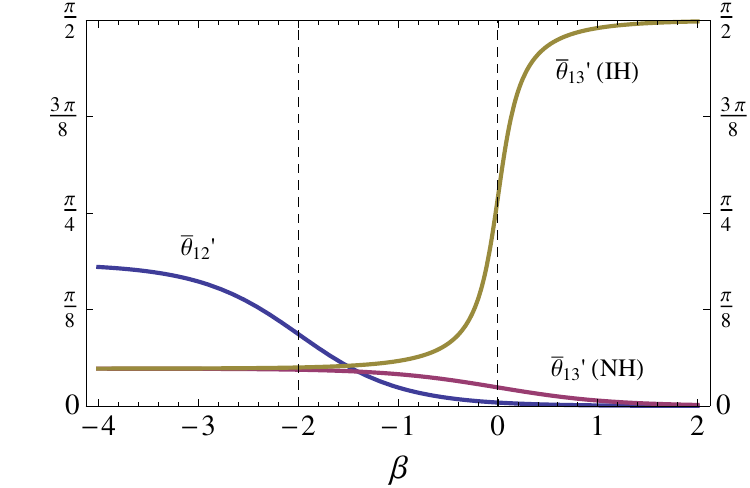}}
\caption{The dependences of the effective mixing angles on $\beta=-\log_\varepsilon(a/|\delta m^2_{31}|)$
for the neutrino (a) and antineutrino (b) cases. 
$\beta=0$ corresponds to $a=|\delta m^2_{31}|$, and $\beta=-2$
to $a=\delta m^2_{21}$.
The $\beta$-dependences of $\theta'_{13}$ and $\overline{\theta}'_{13}$ depend on the mass
hierarchy: when $\delta m^2_{31}>0$ (normal hierarchy, NH) $\theta'_{13}$ increases toward
$\pi/2$ whereas $\overline{\theta}'_{13}$ decreases toward zero,
while in the $\delta m^2_{31}<0$ case (inverted hierarchy, IH), it is the other way around.}
\label{thetaprimes}
\end{center}
\end{figure}
%
\begin{figure}
\begin{center}
\subfigure[neutrino case]{\includegraphics[height=5cm]{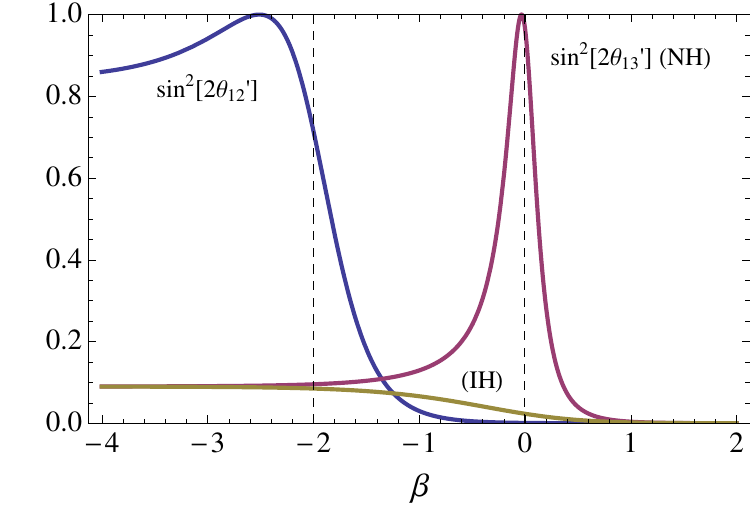}}
\subfigure[anti-neutrino case]{\includegraphics[height=5cm]{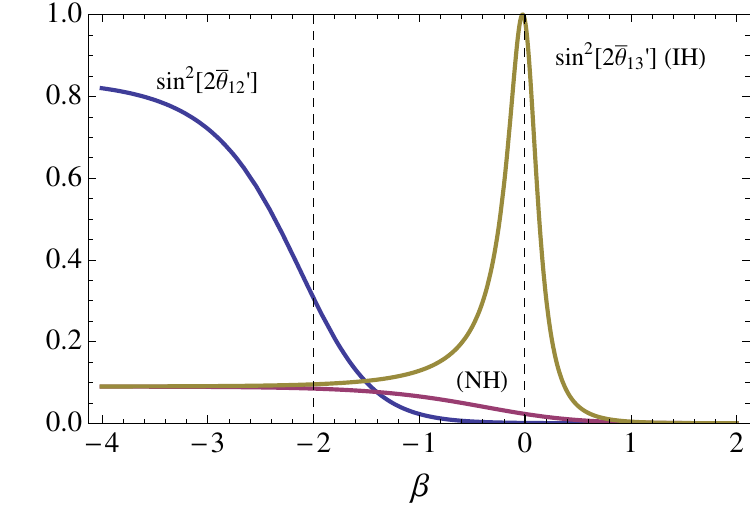}}
\caption{The $\beta$-dependences of the sines of twice the effective mixing angles for the neutrino (a) and antineutrino (b) cases. 
The difference in the behavior of the effective $\theta_{13}$ mixing angle
for normal and inverted hierarchies will allow us to determine which is
chosen by nature.
}
\label{sin2thetas}
\end{center}
\end{figure}
%
\begin{figure}
\subfigure[neutrino, normal hierarchy]{\includegraphics[height=5cm]{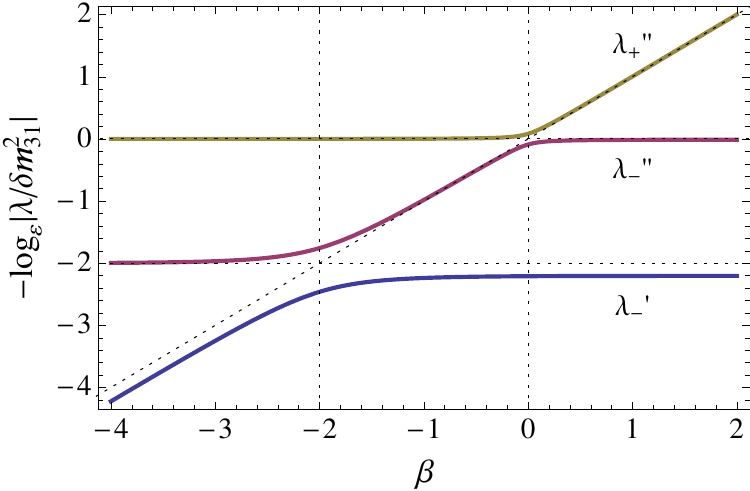}}
\subfigure[neutrino, inverted hierarchy]{\includegraphics[height=5cm]{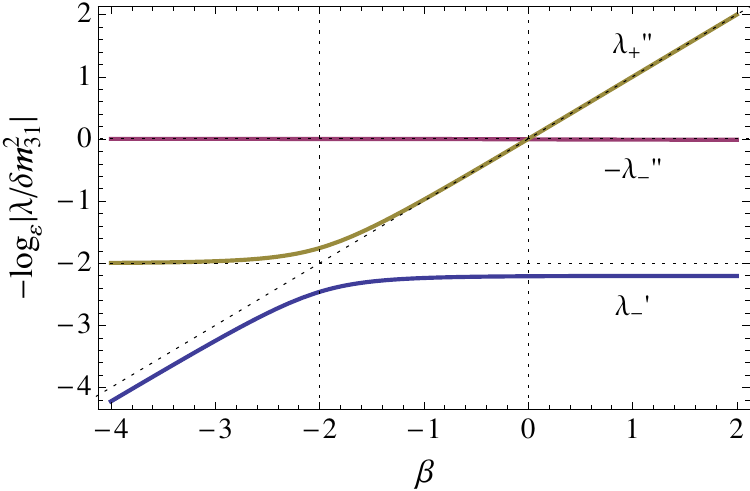}}
\\
\subfigure[anti-neutrino, normal hierarchy]{\includegraphics[height=5cm]{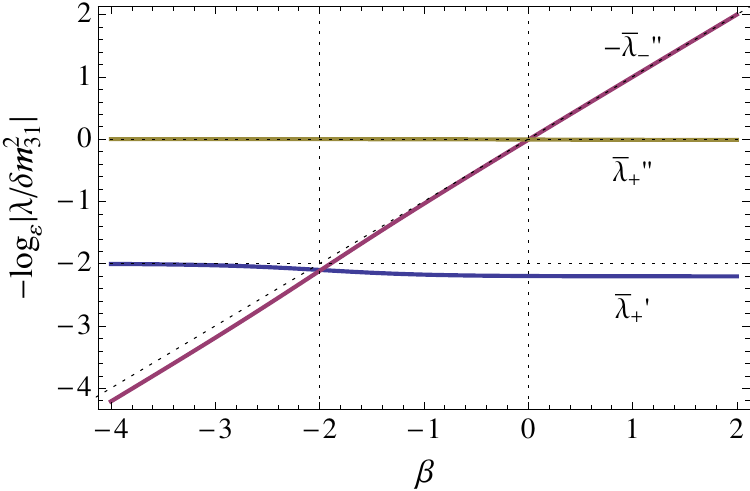}}
\subfigure[anti-neutrino, inverted hierarchy]{\includegraphics[height=5cm]{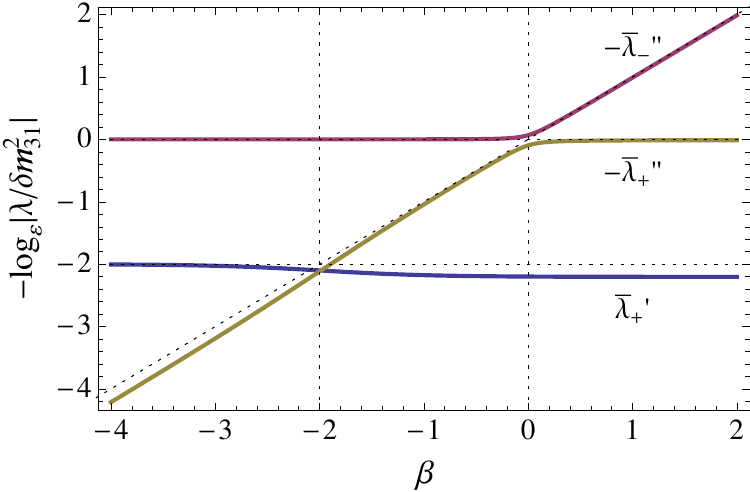}}
\caption{Dependence of the approximate eigenvalues of the effective Hamiltonian on 
$\beta=-\log_\varepsilon(a/|\delta m^2_{31}|)$ for
the (a) neutrino normal hierarchy, (b) neutrino inverted hierarchy, 
(c) anti-neutrino normal hierarchy, and (d) anti-neutrino inverted hierarchy cases.
}
\label{lambdadoubleprimeplot}
\end{figure}

\subsection{The $\beta$-dependence of Mixing Parameters}
\label{subsec:beta_definition}
%
We show plots depicting how our various
effective parameters run with the matter-effect parameter $a$.  
Due to the wide separation in scale
between $\delta m^2_{21}$ and $\delta m^2_{31}$,
we find it convenient to introduce the parameter $\beta$ via\footnote{%
We avoid the use of the symbols $\alpha$ or $A$ since they often
respectively denote $\delta m^2_{21}/\delta m^2_{31}$ and $a/\delta m^2_{31}$ in the literature.}
\begin{equation}
\dfrac{a}{|\delta m^2_{31}|}\;=\;\varepsilon^{-\beta}
\;,\qquad
\varepsilon\;\equiv\;\sqrt{\dfrac{\delta m^2_{21}}{|\delta m^2_{31}|}}
\;\approx\; 0.17
\;,
\label{betadef}
\end{equation}
and plot our effective running parameters as functions of $\beta$ instead of $a$.
Here $\beta=0$ corresponds to $a=|\delta m^2_{31}|$,
$\beta=-2$ to $a=\delta m^2_{21}$, 
and so on.
The dependence of the effective mixing angles on $\beta$ are shown in Fig.~\ref{thetaprimes} and that of the sines of twice these angles in Fig.~\ref{sin2thetas}.
The $\beta$-dependence of approximate eigenvalues of the effective Hamiltonian are shown in Fig.~\ref{lambdadoubleprimeplot}.
%
%

\section{Demonstration of the Accuracy of the Approximation}

In this section, we plot neutrino oscillation probabilities in several scenarios to demonstrate the accuracy of our approximation.
As seen in the previous section, our formulae for both the neutrino and
anti-neutrino cases are fairly compact and easy to code.
In particular, the effective mixing angles for the
neutrino and anti-neutrino cases can be calculated with the
same code by simply flipping the sign of the matter-effect parameter $a$,
\textit{cf.} Eqs.~(\ref{effectivethetadef}) and (\ref{effectivethetabardef}).
The same can be said of $\lambda'_\pm$ and $\overline{\lambda}'_\pm$
defined in Eqs.~(\ref{lambdaprimesdef}) and (\ref{lambdabarprimesdef}).
In the case of $\lambda''_\pm$ and $\overline{\lambda}''_\pm$,
one also needs to make the swap $\lambda'_+ \leftrightarrow \overline{\lambda}_-$ but otherwise the code will be essentially the same.
For the vacuum values of the mixing angles and mass-squared differences, we use
the global fit values from Ref.~\cite{GonzalezGarcia:2012sz} listed in Table~\ref{tab:bench}. All plots are generated assuming constant Earth matter density.

\begin{table}[b]
\begin{center}
\begin{tabular}{|c|c|} 
\hline
$\quad\delta m^2_{21}\quad$ & $\quad 7.5\phantom{0} \times 10^{-5} \ {\rm eV}^2\quad$  \\
\hline
$\delta m^2_{31}$ & $2.47 \times 10^{-3} \ {\rm eV}^2$ \\
\hline
$\sin^2\theta_{23}$ & $0.5\phantom{00}$ \\
\hline
$\sin^2\theta_{12}$ & $0.3\phantom{00}$ \\
\hline
$\sin^2\theta_{13}$ & $0.023$ \\
\hline
\end{tabular}
\caption{Best-fit values of oscillation parameters taken from 
Ref.~\cite{GonzalezGarcia:2012sz}.}
\label{tab:bench}
\end{center}
\end{table}

\begin{figure}[t]
\subfigure[]{\includegraphics[width=7.5cm]{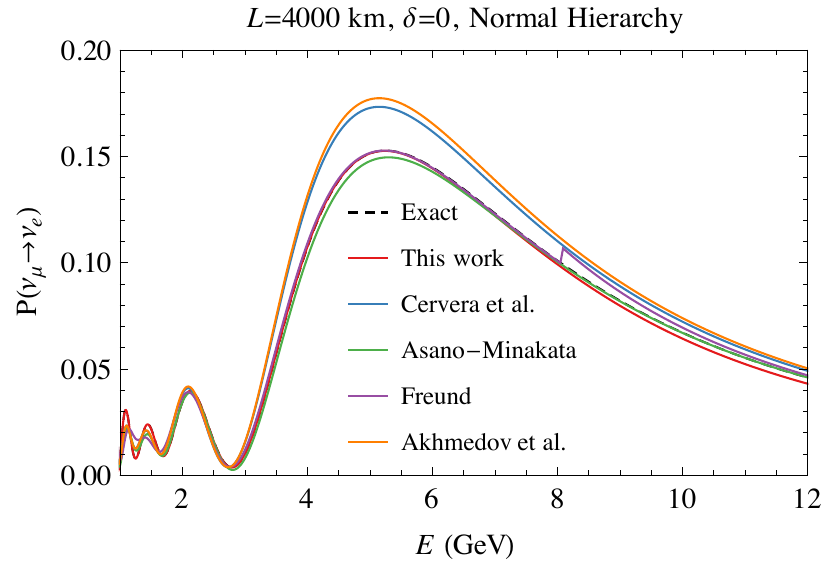}}
\subfigure[]{\includegraphics[width=7.5cm]{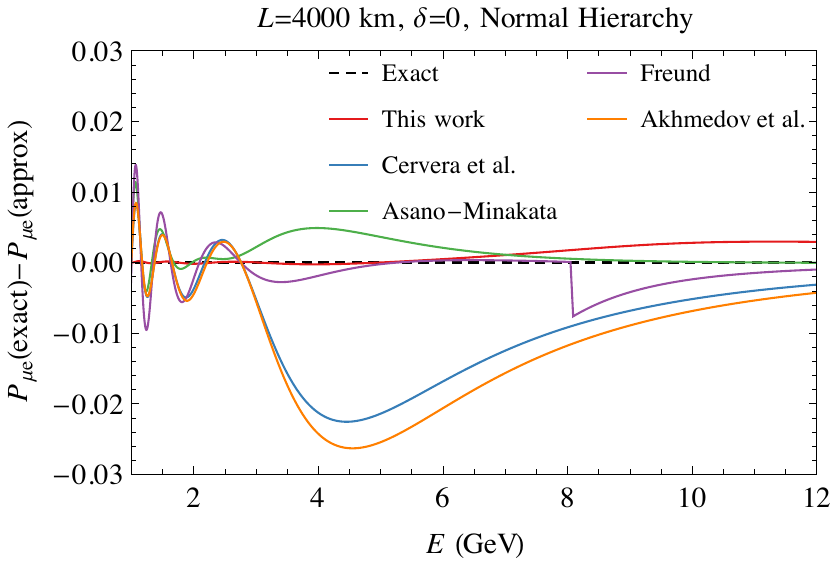}}
\caption{\label{fig:4000km} Comparison of the approximation formulae of
Cervera et al., Akhmedov et al., Asano-Minakata, Freund, and this work at $L=4000\,\mathrm{km}$.
In left panel, the dashed line gives the exact numerical result assuming the line-averaged constant matter density 
of $\rho=3.81\,\mathrm{g/cm^3}$. This has been estimated using the PREM profile of the Earth \cite{PREM:1981}. 
}
\end{figure}
%
\begin{figure}[t]
\subfigure[]{\includegraphics[width=7.5cm]{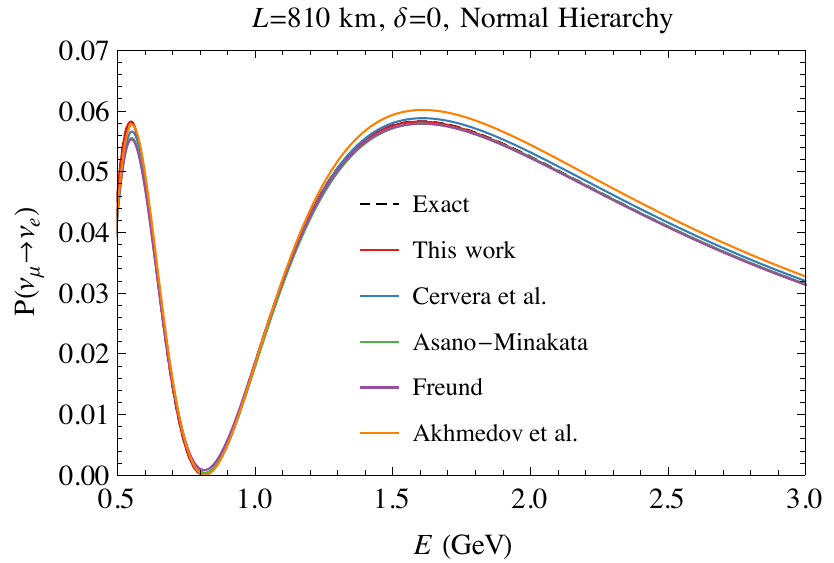}}
\subfigure[]{\includegraphics[width=7.5cm]{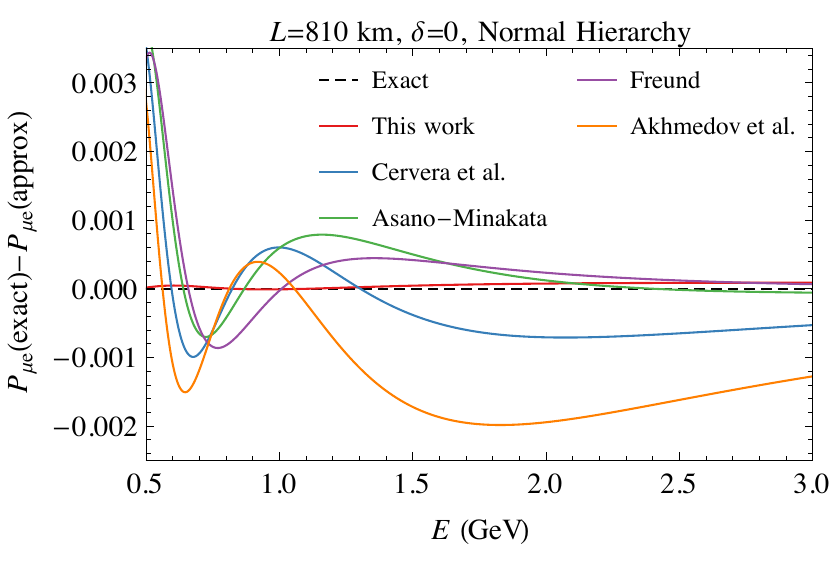}}
\caption{\label{fig:810km} Comparison of the approximation formulae of
Cervera et al., Akhmedov et al., Asano-Minakata, Freund, and this work at $L=810\,\mathrm{km}$, which is
the distance from Fermilab to NO$\nu$A. In left panel, the dashed line gives the exact numerical result
assuming the line-averaged constant matter density of $\rho=2.80\,\mathrm{g/cm^3}$.
This has been estimated using the PREM profile of the Earth \cite{PREM:1981}.
}
\end{figure}
%
\begin{figure}[t]
\subfigure[]{\includegraphics[width=7.5cm]{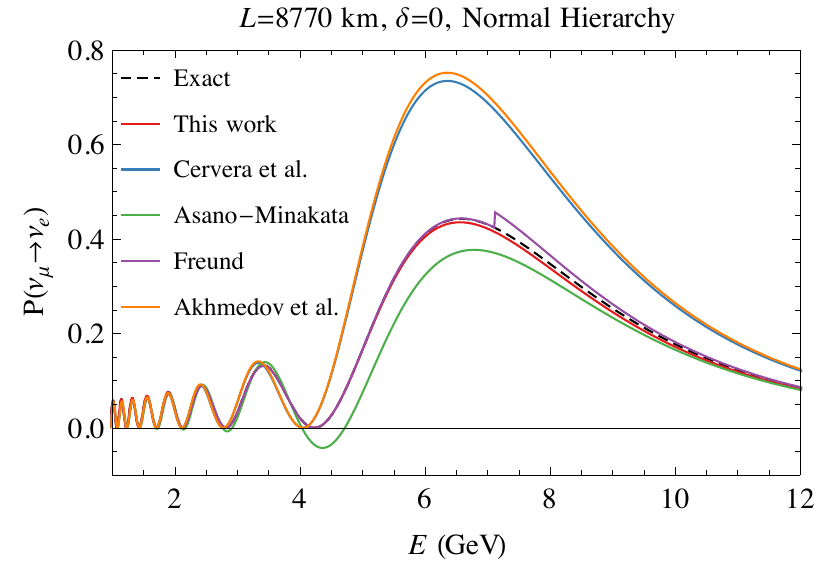}}
\subfigure[]{\includegraphics[width=7.5cm]{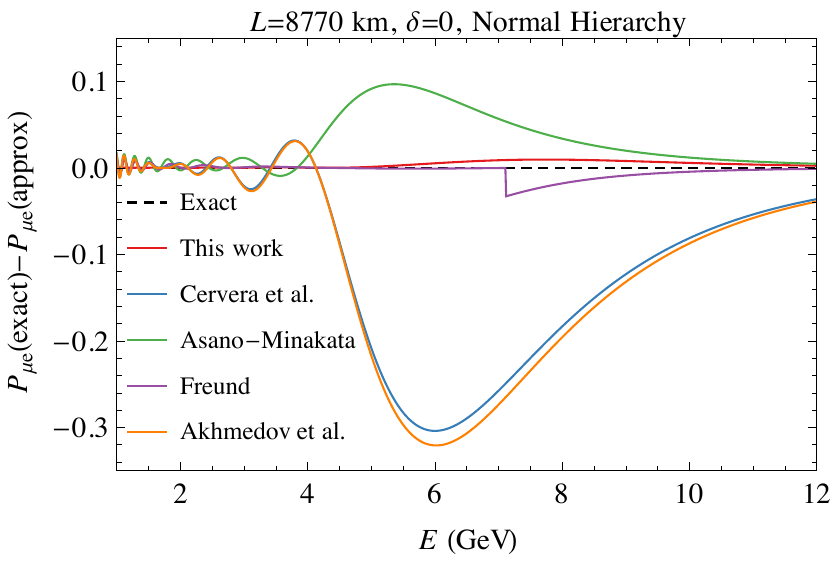}}
\caption{\label{fig:8770km} Comparison of the approximation formulae of
Cervera et al., Akhmedov et al., Asano-Minakata, Freund, and this work at $L=8770\,\mathrm{km}$, which is the distance from 
CERN to Kamioka. In left panel, the dashed line gives the exact numerical result
assuming the line-averaged constant matter density of $\rho=4.33\,\mathrm{g/cm^3}$.
This has been estimated using the PREM profile of the Earth \cite{PREM:1981}.
Note that the Asano-Minakata formula gives negative probability for $E\sim 4\,\mathrm{GeV}$.
}
\end{figure}

\begin{figure}[h]
\subfigure[]{\includegraphics[width=7.5cm]{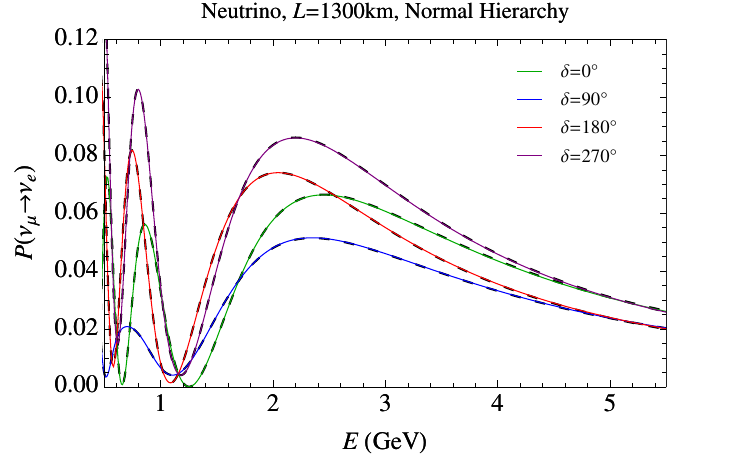}}
\subfigure[]{\includegraphics[width=7.5cm]{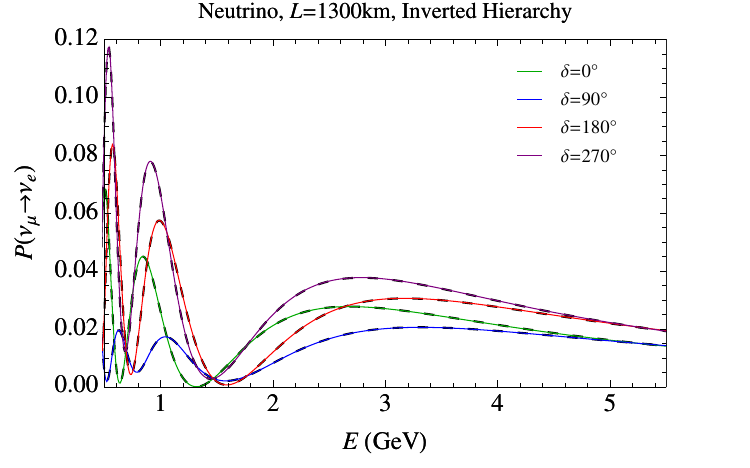}}
\subfigure[]{\includegraphics[width=7.5cm]{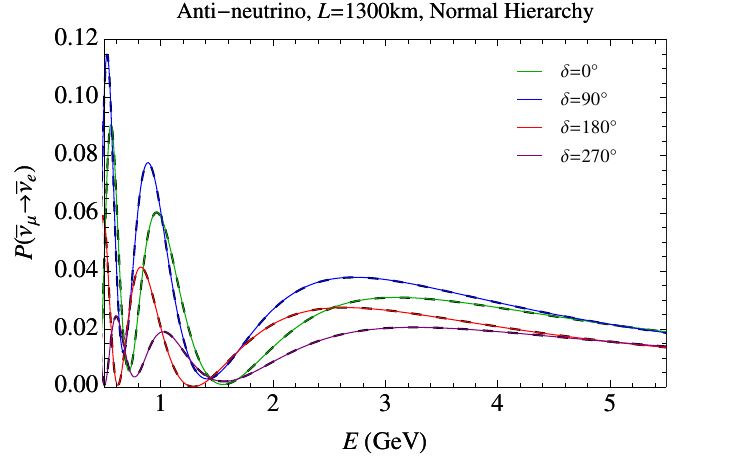}}
\subfigure[]{\includegraphics[width=7.5cm]{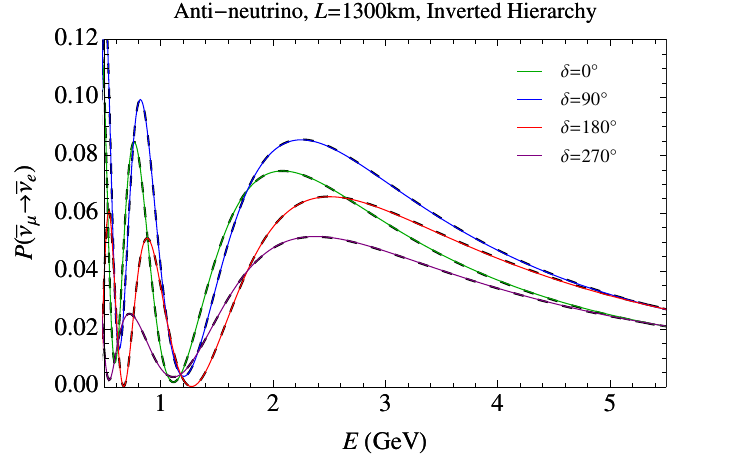}}
\caption{
Comparison of our approximation formulae (colored) to the exact numerical results (black, dashed)
for various values of the CP violating phase $\delta$ at $L=1300\,\mathrm{km}$.
The line-averaged constant matter density for this baseline length is $\rho=2.87\,\mathrm{g/cm^3}$. 
}
\label{fig:CP-check-1300}
\end{figure}
%
\begin{figure}[h]
\subfigure[]{\includegraphics[width=8cm]{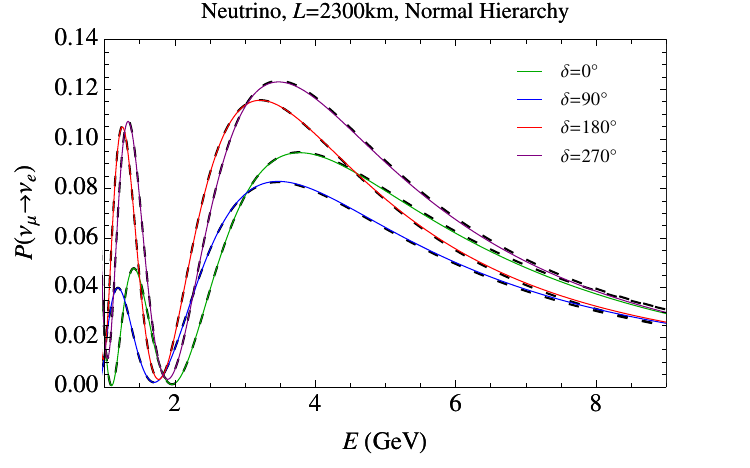}}
\subfigure[]{\includegraphics[width=8cm]{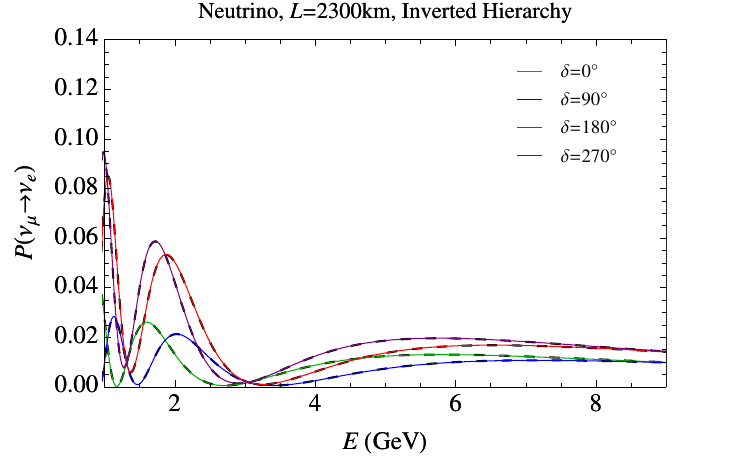}}
\subfigure[]{\includegraphics[width=8cm]{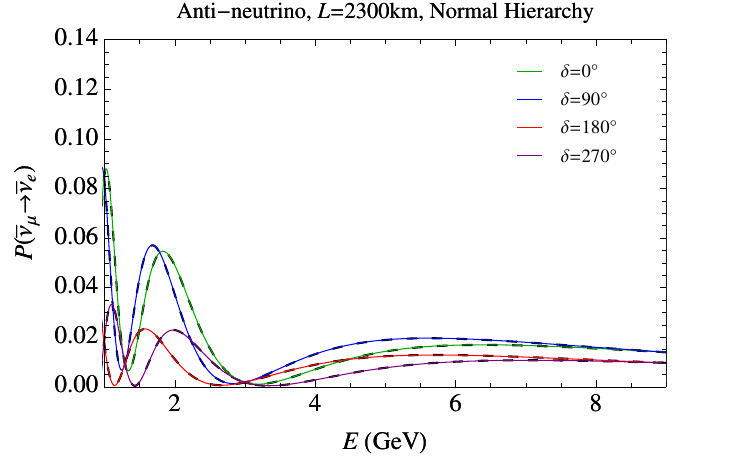}}
\subfigure[]{\includegraphics[width=8cm]{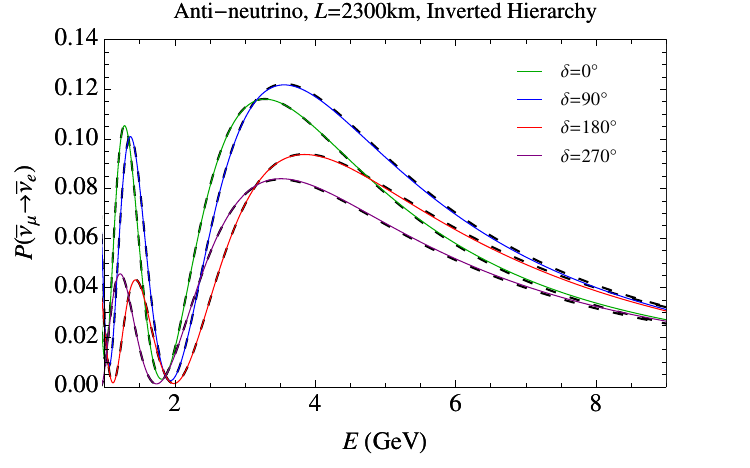}}
\caption{
Comparison of our approximation formulae (colored) to the exact numerical results (black, dashed)
for various values of the CP violating phase $\delta$ at $L=2300\,\mathrm{km}$.
The line-averaged constant matter density for $L=2300\,\mathrm{km}$ is $\rho=3.54\,\mathrm{g/cm^3}$. 
}
\label{fig:CP-check-2300}
\end{figure}

We begin by comparing our approximation to 
Eq.~(16) of Cervera et al. \cite{Cervera:2000kp},
Eq.~(3.5) of Akhmedov et al. \cite{Akhmedov:2004ny},
sum of Eqs.~(4.2) to (4.4) of Asano and Minakata \cite{Asano:2011nj}, and Eq.~(36) of Freund \cite{Freund:2001pn}.
Note that both Cervera et al. and Akhmedov et al. expand the oscillation probabilities to the
same order, so their expressions are quite similar except for a minor difference:
Eq.~(16) of Cervera et al is the same as Eq.~(38) of Freund,
while Eq.~(3.5) of Akhmedov et al. is obtained from 
the same by setting $\cos\theta_{13}=1$
while keeping $\sin\theta_{13}$ non-zero.

In Fig.~\ref{fig:4000km}(a), we plot the approximate $\nu_\mu\rightarrow\nu_e$ oscillation
probabilities calculated using these three approximations against the exact numerical result
for the baseline length $L=4000\,\mathrm{km}$.
This is the distance used by Asano and Minakata in Ref.~\cite{Asano:2011nj} to
demonstrate the strength of their formula.
The line-averaged constant Earth matter density\footnote{All the results presented in this paper have been derived
assuming the line-averaged constant Earth matter density (based on the PREM profile) for a given baseline.}
for this baseline is $3.81\mathrm{g/cm^3}$ which has been 
estimated using the Preliminary Reference Earth Model (PREM) \cite{PREM:1981}.
We consider the normal hierarchy case, $\delta m^2_{31}>0$, with the CP violating phase $\delta$ set to zero.
The differences between the exact and approximate formulae are plotted in Fig.~\ref{fig:4000km}(b).
As can be seen, at this baseline, both the Asano-Minakata formula and our approximation work
much better than the Cervera et al. or the Akhmedov et al. formulae. 
The Freund formula works well in the energy range $E\alt 8\,\mathrm{GeV}$,
but leads to a kink at $E\sim 8\,\mathrm{GeV}$ due to some terms
in the expression changing sign at $a=|\delta m_{13}^2|\cos 2\theta_{13}$.  

The comparison at a shorter baseline length of $L=810\,\mathrm{km}$, which is the distance from
Fermilab to NO$\nu$A, is shown in Fig.~\ref{fig:810km}. There, all five approximations work well, with our 
approximation being the most accurate.

The situation changes at the longer baseline length of $L=8770\,\mathrm{km}$, which is the distance
from CERN to Kamioka \cite{Agarwalla:2012zu}, as can be seen in Fig.~\ref{fig:8770km}.
There, the Cervera et al. and the Akhmedov et al. formulae greatly overestimate $P(\nu_\mu\rightarrow\nu_e)$, 
while the Asano-Minakata formula leads to negative probability for $E\sim 4\,\mathrm{GeV}$.
The Freund formula is accurate up until $E\sim 7\,\mathrm{GeV}$ where a kink occurs at $a=|\delta m_{13}^2|\cos 2\theta_{13}$.
In comparison, our approximation remains accurate for all energies.


The accuracy of our approximation for both the neutrino and anti-neutrino cases,
and both mass hierarchies, for different values of the CP violating phase $\delta$, is demonstrated in 
Figs.~\ref{fig:CP-check-1300} and \ref{fig:CP-check-2300} for the
two baselines $L=1300\,\mathrm{km}$ and $L=2300\,\mathrm{km}$, respectively.
These distances correspond to those between Fermilab and Homestake (1300 km),
and CERN and Pyh\"asalmi (2300 km) \cite{Stahl:2012exa}.
As is evident, our approximation maintains its accuracy for all energy ranges and mass densities.

\section{Applications}
\label{sec:applications}

\subsection{Determination of the Mass Hierarchy from $\nu_e$ Oscillations}
\label{subsec:mass-hierarchy-e}

Consider the $\nu_e$ survival probability in matter which is given by
\begin{eqnarray}
\lefteqn{P(\nu_e\rightarrow\nu_e)}
\cr
& = & 1 - 4\, |\tilde{U}_{e 2}|^2 \left( 1 - |\tilde{U}_{e 2}|^2 \right)
            \sin^2\frac{\tilde{\Delta}_{21}}{2}
        - 4\, |\tilde{U}_{e 3}|^2 \left( 1 - |\tilde{U}_{e 3}|^2 \right)
            \sin^2\frac{\tilde{\Delta}_{31}}{2} 
\cr
& &  \phantom{1}
        + 2\, |\tilde{U}_{e 2}|^2 |\tilde{U}_{e 3}|^2
          \left( 4\sin^2\frac{\tilde{\Delta}_{21}}{2}\sin^2\frac{\tilde{\Delta}_{31}}{2}
                + \sin\tilde{\Delta}_{21}\sin\tilde{\Delta}_{31}
          \right) 
\cr
& = & 1 - 4\,c_{13}^{\prime 2} s_{12}^{\prime 2}\left(1-c_{13}^{\prime 2} s_{12}^{\prime 2}\right)
            \sin^2\frac{\tilde{\Delta}_{21}}{2}
        - \sin^2(2\theta'_{13})\sin^2\frac{\tilde{\Delta}_{31}}{2}
\cr
& &  \phantom{1}
        + s_{12}^{\prime 2} \sin^2(2\theta'_{13})
          \left( 2\sin^2\frac{\tilde{\Delta}_{21}}{2}\sin^2\frac{\tilde{\Delta}_{31}}{2}
                + \dfrac{1}{2}\sin\tilde{\Delta}_{21}\sin\tilde{\Delta}_{31}
          \right) 
\cr
& \xrightarrow{s'_{12}\approx 1} & 1
- \sin^2(2\theta'_{13})
\Biggl(
\sin^2\frac{\tilde{\Delta}_{21}}{2}+\sin^2\frac{\tilde{\Delta}_{31}}{2}
- 2\sin^2\frac{\tilde{\Delta}_{21}}{2}\sin^2\frac{\tilde{\Delta}_{31}}{2}
- \dfrac{1}{2}\sin\tilde{\Delta}_{21}\sin\tilde{\Delta}_{31}
\Biggr) 
\cr
& = & 1 - \sin^2(2\theta'_{13})\;\sin^2\dfrac{\tilde{\Delta}_{32}}{2}
\;,
\label{Pee}
\end{eqnarray}
where we have assumed that $a\gg \delta m^2_{21}$ so that $s'_{12}\approx 1$ is a good
approximation.
Similarly, we find:
\begin{eqnarray}
\lefteqn{P(\nu_e \rightarrow \nu_\mu)} \cr
& = & 4\, |\tilde{U}_{e 2}|^2 |\tilde{U}_{\mu 2}|^2 \sin^2\frac{\tilde{\Delta}_{21}}{2}
     +4\, |\tilde{U}_{e 3}|^2 |\tilde{U}_{\mu 3}|^2 \sin^2\frac{\tilde{\Delta}_{31}}{2} \cr
&   & +2\;\Re\left( \tilde{U}^*_{e 3}\tilde{U}_{\mu 3}\tilde{U}_{e 2}\tilde{U}^*_{\mu 2}\right)
      \left(4\sin^2\frac{\tilde{\Delta}_{21}}{2}\sin^2\frac{\tilde{\Delta}_{31}}{2}
           +\sin\tilde{\Delta}_{21}\sin\tilde{\Delta}_{31}
      \right) \cr 
&   & +4\,\tilde{J}_{(e,\mu)}
      \left( \sin^2\frac{\tilde{\Delta}_{21}}{2}\sin\tilde{\Delta}_{31}
            -\sin^2\frac{\tilde{\Delta}_{31}}{2}\sin\tilde{\Delta}_{21}
      \right)   
\label{Pe2mu1}\\
& = & 4\, s^{\prime 2}_{12}c^{\prime 2}_{13} 
      \left(
      c^{\prime 2}_{12}c^2_{23}+s^{\prime 2}_{12}s^{\prime 2}_{13}s^2_{23}
      -2s'_{12}c'_{12}s'_{13}c_{23}s_{23}\cos\delta
      \right) 
      \sin^2\frac{\tilde{\Delta}_{21}}{2}
     +4\, s^{\prime 2}_{13} c^{\prime 2}_{13} s^2_{23} \sin^2\frac{\tilde{\Delta}_{31}}{2} \cr
&   & +2\;s'_{12}s'_{13}c^{\prime 2}_{13}s_{23}
      \left(c'_{12}c_{23}\cos\delta - s'_{12}s'_{13}s_{23}\right)
      \left(4\sin^2\frac{\tilde{\Delta}_{21}}{2}\sin^2\frac{\tilde{\Delta}_{31}}{2}
           +\sin\tilde{\Delta}_{21}\sin\tilde{\Delta}_{31}
      \right) \cr 
&   & -4\,s'_{12}c'_{12}s'_{13}c^{\prime 2}_{13}s_{23}c_{23}\sin\delta
      \left( \sin^2\frac{\tilde{\Delta}_{21}}{2}\sin\tilde{\Delta}_{31}
            -\sin^2\frac{\tilde{\Delta}_{31}}{2}\sin\tilde{\Delta}_{21}
      \right)     
\label{Pe2mu2}\\
& \xrightarrow{s'_{12}\approx 1} &
s^2_{23}\,\sin^2(2\theta'_{13})\;\sin^2\dfrac{\tilde{\Delta}_{32}}{2}
\;,
\label{Pe2mu3}
\\
& & \cr
%
\lefteqn{P(\nu_e \rightarrow \nu_\tau)} \cr
& = & 4\, |\tilde{U}_{e 2}|^2 |\tilde{U}_{\tau 2}|^2 \sin^2\frac{\tilde{\Delta}_{21}}{2}
     +4\, |\tilde{U}_{e 3}|^2 |\tilde{U}_{\tau 3}|^2 \sin^2\frac{\tilde{\Delta}_{31}}{2} \cr
&   & +2\;\Re\left( \tilde{U}^*_{e 3}\tilde{U}_{\tau 3}\tilde{U}_{e 2}\tilde{U}^*_{\tau 2}\right)
      \left(4\sin^2\frac{\tilde{\Delta}_{21}}{2}\sin^2\frac{\tilde{\Delta}_{31}}{2}
           +\sin\tilde{\Delta}_{21}\sin\tilde{\Delta}_{31}
      \right) \cr 
&   & +4\,\tilde{J}_{(e,\tau)}
      \left( \sin^2\frac{\tilde{\Delta}_{21}}{2}\sin\tilde{\Delta}_{31}
            -\sin^2\frac{\tilde{\Delta}_{31}}{2}\sin\tilde{\Delta}_{21}
      \right) 
\cr
& = & 4\, s^{\prime 2}_{12}c^{\prime 2}_{13} 
      \left(
      c^{\prime 2}_{12}s^2_{23}+s^{\prime 2}_{12}s^{\prime 2}_{13}c^2_{23}
      -2s'_{12}c'_{12}s'_{13}s_{23}c_{23}\cos\delta
      \right) 
      \sin^2\frac{\tilde{\Delta}_{21}}{2}
     +4\, s^{\prime 2}_{13} c^{\prime 2}_{13}c^2_{23} \sin^2\frac{\tilde{\Delta}_{31}}{2} \cr
&   & -2\;s'_{12}s'_{13}c^{\prime 2}_{13}c_{23}
      \left(
      c'_{12}s_{23}\cos\delta
      + s'_{12}s'_{13}c_{23}
      \right)
      \left(4\sin^2\frac{\tilde{\Delta}_{21}}{2}\sin^2\frac{\tilde{\Delta}_{31}}{2}
           +\sin\tilde{\Delta}_{21}\sin\tilde{\Delta}_{31}
      \right) \cr 
&   & +4\,s'_{12}c'_{12}s'_{13}c^{\prime 2}_{13}s_{23}c_{23}\sin\delta
      \left( \sin^2\frac{\tilde{\Delta}_{21}}{2}\sin\tilde{\Delta}_{31}
            -\sin^2\frac{\tilde{\Delta}_{31}}{2}\sin\tilde{\Delta}_{21}
      \right) 
\cr
& \xrightarrow{s'_{12}\approx 1} &
c^2_{23}\,\sin^2(2\theta'_{13})\;\sin^2\dfrac{\tilde{\Delta}_{32}}{2}
\;.
\label{Pe2tau}       
\end{eqnarray}
From Fig.~\ref{sin2thetas}, it is clear that the factor
$\sin^2(2\theta'_{13})$ in these expressions
behaves quite differently depending on the mass hierarchy.
For normal hierarchy $\sin^2(2\theta'_{13})$ will peak around 
$a\approx \delta m^2_{31}$ but for the inverted hierarchy case it will not.
This will become manifest if the factor $\sin^2(\tilde{\Delta}_{32}/2)$
also peaked at or near the same energy.\footnote{%
If we expand the running parameters in our Eq.~(\ref{Pe2mu3}) in powers of 
the vacuum $s_{13}$ and $\alpha=\delta m^2_{21}/\delta m^2_{31}$, 
the leading order term expressed using the notation of Freund \cite{Freund:2001pn} 
takes the form
\[
P(\nu_e \rightarrow \nu_\mu)
\;=\;
s^2_{23}\,\sin^2(2\theta'_{13})\;\sin^2\dfrac{\tilde{\Delta}_{32}}{2}
\;=\; s^2_{23}\,\dfrac{4s^2_{13}}{(\hat{A}-1)^2}\sin^2\left[(\hat{A}-1)\hat{\Delta}\right]
+ \cdots
\]
where $\hat{A}=a/\delta m^2_{31}$ and $\hat{\Delta}=\delta m^2_{31}L/4E$.
This is the same as Eq.~(38a) of Freund with $\sin^2 2\theta_{13}$ replaced by $4s_{13}^2$,
and agrees with the corresponding term of Akhmedov et al. \cite{Akhmedov:2004ny}.
The enhancement discussed in the main text can be seen to occur at $\hat{A}=1$, that is $a=\delta m^2_{31}$, which is possible only when $\delta m^2_{31}>0$.
However, the formulae 
of Cervera et al. \cite{Cervera:2000kp},
Akhmedov et al. \cite{Akhmedov:2004ny},
Asano and Minakata \cite{Asano:2011nj}, and Freund \cite{Freund:2001pn}
compared in the previous section all suffer in accuracy around the resonance region $\hat{A}\approx 1$.
This is not the case for our expression, which has a smooth transition across the resonance.
The fact that the CP-phase dependent terms are negligible at the relevant energies and baselines is
also clear in our approach.
}

\begin{figure}[t]
\subfigure[]{\includegraphics[width=7.5cm]{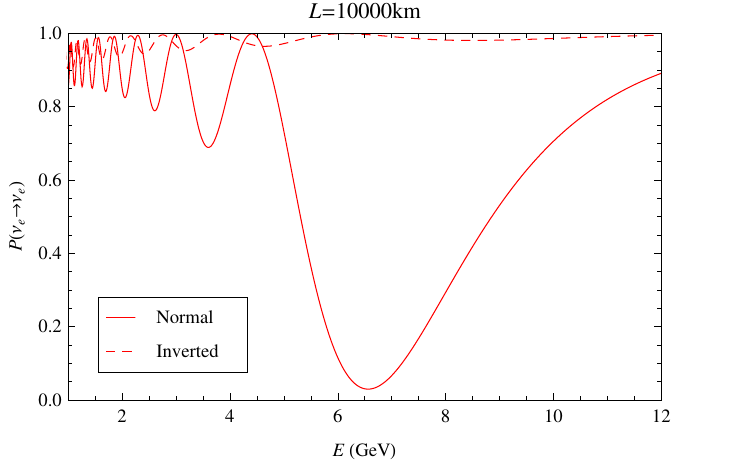}}
\subfigure[]{\includegraphics[width=7.5cm]{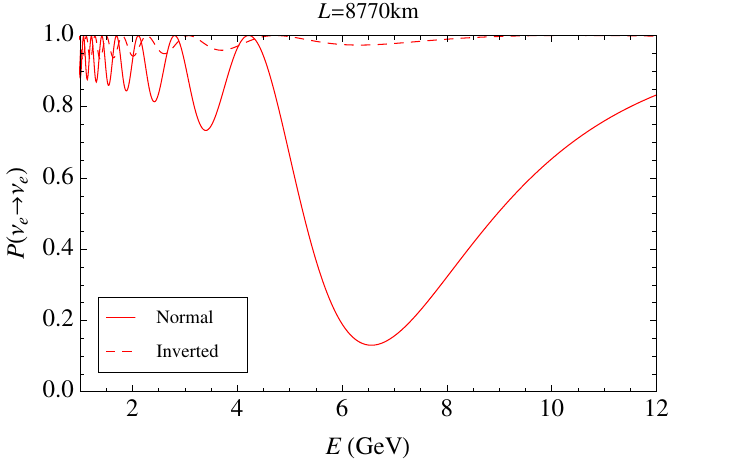}}
\caption{\label{fig:NIcomparison} Comparison of the exact
oscillation probabilities $P(\nu_e\rightarrow\nu_e)$
between the normal and inverted hierarchies at
(a) $L=10000\,\mathrm{km}$ ($\rho=4.53\,\mathrm{g/cm^3}$), and
(b) $L=8770\,\mathrm{km}$ ($\rho=4.33\,\mathrm{g/cm^3}$)
}
\end{figure}

For the normal hierarchy case, when $a\approx\delta m^2_{31}$ we have
\begin{equation}
\delta\lambda_{32}
\;=\; \lambda''_+-\lambda''_-
\;\approx\; \sqrt{[\lambda'_+ - (\delta m^2_{31}+a s^2_{13})]^2
+ 4a^2 c^2_{13}s^2_{13}}
\;\approx\;
2 s_{13}\,a
\;.
\end{equation}
Therefore,
\begin{eqnarray}
\dfrac{\tilde{\Delta}_{32}}{2}
\;=\; \dfrac{\delta\lambda_{32}}{4E}L
\;\approx\; \dfrac{s_{13}\,a}{2E}L
& = &
\left(2.9\times 10^{-5}\right)
\left(\dfrac{\rho}{\mathrm{g/cm^3}}\right)
\left(\dfrac{L}{\mathrm{km}}\right)
\cr
& = & \dfrac{\pi}{2}
\left(\dfrac{\rho L}{54000\;\mathrm{(km\cdot g/cm^3)}}\right)
\;.
\end{eqnarray}
From Fig.~\ref{rhofig}, it is clear that 
$\rho L < 54000\;\mathrm{(km\cdot g/cm^3)}$ as long as the neutrino beam does not
enter the core of the Earth, at which point the constant average matter density approximation
breaks down.
Therefore, in order to take $\tilde{\Delta}_{32}/2$ as close as possible to 
$\pi/2$ while preventing the beam from entering the Earth's core, we need
$L\sim 10000\,\mathrm{km}$.

For instance, if we take $L=10000\,\mathrm{km}$
for which $\rho=4.53\,\mathrm{g/cm^3}$, we have
$\rho L \approx 45300\,\mathrm{km\cdot g/cm^3}$.
The value of $\tilde{\Delta}_{32}/2$ at resonance $a\approx\delta m^2_{31}$ is then
\begin{equation}
\dfrac{\pi}{2}\times\dfrac{45300}{54000} \;=\; 0.42\,\pi\;,
\end{equation}
leading to an oscillation peak/dip factor of $\sin^2(\tilde{\Delta}_{32}/2)=0.94$.
Using Eq.~(\ref{adef}), the neutrino beam energy at which $a\approx\delta m^2_{31}$ is found to be
\begin{equation}
\dfrac{E}{\mathrm{GeV}}
\;=\; \dfrac{(\delta m^2_{31}/\mathrm{eV^2})}{(7.63\times 10^{-5})\times(\rho/(\mathrm{g/cm^3}))}
\;=\; \dfrac{(2.47\times 10^{-3})}{(7.63\times 10^{-5})\times (4.53)}
\;\approx\; 7
\;.
\end{equation}
Indeed, in Fig.~\ref{fig:NIcomparison}(a) we show the exact $\nu_e$ survival probabilities 
at $L=10000\,\mathrm{km}$ for both hierarchies, and we can see that the normal hierarchy case
dips by over 95\% around $E\sim 6.5\,\mathrm{GeV}$.
Thus, our rough estimates give a highly reliable result.

If we take a somewhat shorter baseline of $L=8770\,\mathrm{km}$, 
which is the distance between CERN and Kamioka \cite{Agarwalla:2012zu},
we have $\rho = 4.33\,\mathrm{g/cm^3}$, and 
$\rho L \approx 38000\,\mathrm{km\cdot g/cm^3}$.
The value of $\tilde{\Delta}_{32}/2$ at resonance $a\approx\delta m^2_{31}$ is then
\begin{equation}
\dfrac{\pi}{2}\times\dfrac{38000}{54000} \;=\; 0.35\,\pi\;,
\end{equation}
leading to an oscillation peak/dip factor of $\sin^2(\tilde{\Delta}_{32}/2)=0.8$, which
is still fairly prominent.
Using Eq.~(\ref{adef}), the neutrino beam energy at which $a\approx\delta m^2_{31}$ is found to be
\begin{equation}
\dfrac{E}{\mathrm{GeV}}
\;=\; \dfrac{(\delta m^2_{31}/\mathrm{eV^2})}{(7.63\times 10^{-5})\times(\rho/(\mathrm{g/cm^3}))}
\;=\; \dfrac{(2.47\times 10^{-3})}{(7.63\times 10^{-5})\times (4.33)}
\;\approx\; 7.5
\;.
\end{equation}
The actual oscillation peak occurs slightly off resonance around $E=6.5\,\mathrm{GeV}$ 
as can already be seen in Fig.~\ref{fig:8770km}.
Comparison of $P(\nu_e\rightarrow\nu_e)$ at $L=8770\,\mathrm{km}$ with $\delta=0$
for the normal and inverted hierarchies are shown in Fig.~\ref{fig:NIcomparison}(b).
$P(\nu_e\rightarrow\nu_\mu)$ is compared in Fig.~\ref{fig:Magic}(b).  

The differences between the normal and inverted hierarchies 
for both baselines is manifest, indicating that
measuring these oscillation probabilities at this baseline would allow us to determine the
mass hierarchy easily.  
(We consider the dependence on the CP violating phase $\delta$ in the next section.)
Eqs.~(\ref{Pee}), (\ref{Pe2mu3}), and (\ref{Pe2tau}) also suggest that the measurement
may provide a better determination of $\sin^2\theta_{23}$.

\subsection{The ``Magic'' Baseline}
\label{subsec:magicbaseline}

\begin{figure}[t]
\subfigure[]{\includegraphics[width=7.5cm]{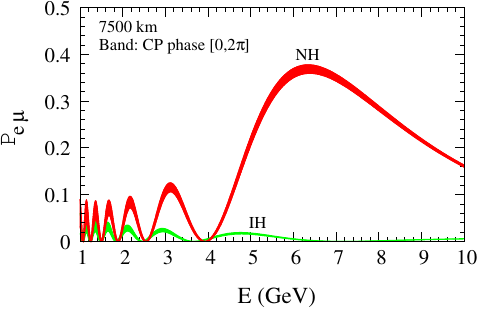}}
\subfigure[]{\includegraphics[width=7.5cm]{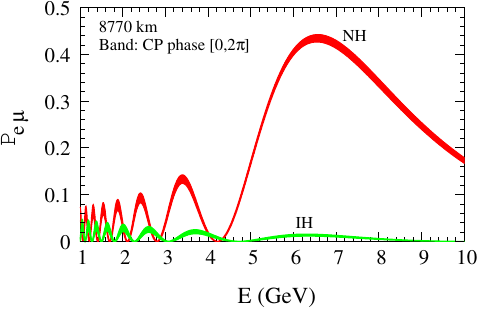}}
\caption{\label{fig:Magic}
The dependence of the exact oscillation probability $P(\nu_e\rightarrow\nu_\mu)$
on the CP violating phase $\delta$ at
(a) $L=7500\,\mathrm{km}$ ($\rho=4.21\,\mathrm{g/cm^3}$), and
(b) $L=8770\,\mathrm{km}$ ($\rho=4.33\,\mathrm{g/cm^3}$)
for the normal (red) and inverted (green) mass hierarchies.
}
\end{figure}

The ``magic'' baseline is the baseline at which the dependence 
of $P(\nu_e\rightarrow\nu_\mu)$ on the CP violating phase $\delta$ vanishes \cite{Huber:2003ak}.\footnote{An illuminating discussion on the physical meaning of the ``magic baseline'' can be found in
Ref.~\cite{Smirnov:2006sm}.}
Looking at Eq.~(\ref{Pe2mu1}),
the only term without $\delta$-dependence is the
$|\tilde{U}_{e 3}|^2 |\tilde{U}_{\mu 3}|^2$ term.
To make every other term vanish, we must have
\begin{equation}
\sin\dfrac{\tilde{\Delta}_{21}}{2} 
\;=\;
\sin\left(\dfrac{\delta\lambda_{21}}{4E}L\right)
\;=\; 0\;.
\end{equation}
Therefore, the magic baseline condition is
\begin{equation}
\dfrac{\delta\lambda_{21}}{4E}L
\;=\; n\pi\;,
\qquad n\in\mathbb{Z}\;.
\end{equation}
If we are in the energy and mass-density range such that
$\delta m^2_{21} < a < |\delta m^2_{31}|$, we can see from Fig.~\ref{lambdadoubleprimeplot}
that $\delta\lambda_{21}\approx a = 2\sqrt{2}G_F N_e E$, 
and the above condition reduces to
\begin{equation}
\sqrt{2}G_F N_e L \;\approx\; 2n\pi\;,
\end{equation}
which is the usual magic baseline condition.
Using Eq.~(\ref{adef}), this condition for the $n=1$ case becomes
\begin{equation}
\dfrac{\tilde{\Delta}_{21}}{2}
\;\approx\;
\dfrac{a}{4E}L
\;=\; (9.7\times 10^{-5})
\left(\dfrac{\rho}{\mathrm{g/cm^3}}\right)
\left(\dfrac{L}{\mathrm{km}}\right)
\;=\; \pi\;,
\end{equation}
that is
\begin{equation}
\dfrac{\rho L}{\mathrm{km\cdot g/cm^3}}
\;\approx\; 32000\;.
\end{equation}
This is satisfied for $L\approx 7500\,\mathrm{km}$ as can be read off of Fig.~\ref{rhofig}.
Indeed, in Fig.~\ref{fig:Magic}(a) we plot the bands that $P(\nu_e\rightarrow\nu_\mu)$
at $L=7500\,\mathrm{km}$ sweeps for both mass hierarchies when $\delta$ is varied throughout its
range of $[0,2\pi]$. 
We can see that the dependence on $\delta$ is very weak.

However, if we look at Eq.~(\ref{Pe2mu2}) carefully, it is clear that all terms
that include the CP violating phase $\delta$ are multiplied by $c'_{12}$ which goes
to zero when $a\gg\delta m^2_{21}$.  Indeed, this was why $\delta$ did not appear
in Eq.~(\ref{Pe2mu3}).  
The condition $a\gg\delta m^2_{21}$ demands
\begin{equation}
\left(\dfrac{\rho}{\mathrm{g/cm^3}}\right)
\left(\dfrac{E}{\mathrm{GeV}}\right)\;\gg\;1\;,
\label{deltaIndependence}
\end{equation}
which is clearly satisfied around the oscillation peak
for the $L=8770\,\mathrm{km}$ case just discussed in the previous section.
Thus, $P(\nu_e\rightarrow\nu_\mu)$ for this baseline is also
only very weakly dependent on $\delta$ as shown in Fig.~\ref{fig:Magic}(b).
We can conclude that, in general, as long as Eq.~(\ref{deltaIndependence})
is satisfied, one does not need to be at a specific ``magic'' baseline
to suppress the $\delta$-dependence of $P(\nu_e\rightarrow\nu_\mu)$.

\section{Summary}
\label{sec:summary}

We have presented a new and simple approximation for calculating the
neutrino oscillation matter effects.
Our approximation was derived utilizing the Jacobi method \cite{Jacobi:1846}, 
and we show in the appendix that at most two rotations are sufficient for
approximate diagonalization of the effective Hamiltonian.
The two rotation angles are absorbed into effective values of
$\theta_{12}$ and $\theta_{13}$.

As explained in detail in the appendix, the approximation works when
$\theta_{13}=O(\varepsilon)$, where $\varepsilon=\sqrt{\delta m^2_{21}/|\delta m^2_{31}|}=0.17$,
a condition which has been shown to be satisfied by Daya Bay \cite{An:2012eh}
and RENO \cite{Ahn:2012nd}.
Our formulae are compact and can easily be coded as well as be manipulated by hand. 
The application of our method to finding the $\nu_e\rightarrow\nu_{\mu},\nu_{\tau}$
resonance conversion condition, and that to the determination of the 
`magic' baseline \cite{Huber:2003ak,Smirnov:2006sm} have
been demonstrated.

In this paper, only the matter effect due to 
Standard Model $W$ exchange between the neutrinos and matter was considered. 
New Physics effects which distinguish between neutrino flavor 
would add extra terms to the effective Hamiltonian, which would require
further rotations for diagonalization.
This has been discussed previously in Ref.~\cite{Honda:2006gv}, 
and the potential constraints on New Physics from long baseline neutrino oscillations experiments
in Refs.~\cite{Honda:2006di,Honda:2007yd,Honda:2007wv}.
Updates to these works will be presented in future publications.

\acknowledgments

We would like to thank Minako Honda and Naotoshi Okamura for their contributions to 
the earlier version of this work \cite{Honda:2006hp,Honda:2006gv}.
Helpful communications with W.~Liao, T.~Ohlsson, S.~Petcov, P.~Roy, and K.~Takeuchi are gratefully acknowledged.
SKA was supported by the DST/INSPIRE Research Grant [IFA-PH-12],
Department of Science \& Technology, India.
SKA is also grateful for the support of IFIC-CSIC, University of Valencia, Spain, where 
some initial portions of this work was carried out.
TT was supported by the U.S. Department of Energy, grant number DE-FG05-92ER40677 task A,
and by the World Premier International Research Center Initiative (WPI Initiative), MEXT, Japan.

\newpage
\appendix

\section{Conventions, Notation, and Basic Formulae}
\label{sec:basics}

Here, we collect the basic formulae associated with neutrino oscillation 
in order to fix our notation and conventions.

\subsection{The PMNS Matrix}
\label{subsec:pmns}

Assuming three-generation neutrino mixing, the flavor eigenstates
$\ket{\nu_\alpha}$ $(\alpha=e,\mu,\tau)$ are related to 
the three mass eigenstates $\ket{\nu_j}$ $(j=1,2,3)$ via the 
Pontecorvo-Maki-Nakagawa-Sakata (PMNS) matrix \cite{Pontecorvo:1957vz,Maki:1962mu,Pontecorvo:1967fh}
\begin{equation}
(V_{\mathrm{PMNS}})_{\alpha j} 
\;\equiv\; \braket{\nu_\alpha}{\nu_j}\;,
\end{equation}
that is,
\begin{equation}
\begin{array}{lll}
\ket{\nu_j} 
& =\, {\displaystyle \sum_{\alpha=e,\mu,\tau} \ket{\nu_\alpha}\braket{\nu_\alpha}{\nu_j}}
& =\, {\displaystyle \sum_{\alpha=e,\mu,\tau} (V_{\mathrm{PMNS}})_{\alpha j}\ket{\nu_\alpha}}\;,
\\
\ket{\nu_\alpha}
& =\; {\displaystyle \sum_{j=1,2,3} \,\ket{\nu_j}\braket{\nu_j}{\nu_\alpha}}
& =\; {\displaystyle \sum_{j=1,2,3}\, (V_{\mathrm{PMNS}})^*_{\alpha j}\ket{\nu_j}}\;.
\end{array}
\end{equation}
The standard parametrization is given by  
\begin{equation}
V_\mathrm{PMNS} \;=\; U\mathcal{P} \;,
\end{equation}
with\footnote{%
Cervera et al. in Ref.~\cite{Cervera:2000kp} use a different convention in which
the sign of $\delta$ is flipped.}
\begin{eqnarray}
U 
& = & R_{23}(\theta_{23},0)\;R_{13}(\theta_{13},\delta)\;R_{12}(\theta_{12},0)
\phantom{\bigg|} \cr
& = &
\left[ \begin{array}{ccc} 1 & 0 & 0 \\
                          0 &  c_{23} & s_{23} \\
                          0 & -s_{23} & c_{23}
       \end{array}
\right]
\left[ \begin{array}{ccc} c_{13} & 0 & s_{13} e^{-i\delta} \\
                          0 & 1 & 0 \\
                          -s_{13} e^{i\delta} & 0 & c_{13}
       \end{array}
\right]
\left[ \begin{array}{ccc} c_{12} & s_{12} & 0 \\
                         -s_{12} & c_{12} & 0 \\
                          0 & 0 & 1
       \end{array}
\right] \cr
& = &
\left[ \begin{array}{ccc}
c_{12}c_{13} & s_{12}c_{13} & s_{13} e^{-i\delta} \\
-s_{12}c_{23} - c_{12}s_{13}s_{23}e^{i\delta} &
\phantom{-}c_{12}c_{23} - s_{12}s_{13}s_{23}e^{i\delta} & c_{13}s_{23} \\
\phantom{-}s_{12}s_{23} - c_{12}s_{13}c_{23}e^{i\delta} &
-c_{12}s_{23} - s_{12}s_{13}c_{23}e^{i\delta} & c_{13}c_{23}
\end{array} \right] \;,\cr
& & \cr
\mathcal{P}
& = &
\mathrm{diag}(1,e^{i\alpha_{21}/2},e^{i\alpha_{31}/2}) \;.
\phantom{\Bigg|}
\label{UPparam}
\end{eqnarray}
Here, $R_{ij}(\theta,\delta)$ denotes a rotation matrix in the $ij$-plane
of clockwise rotation angle $\theta$ 
with phases $\pm\delta$ on the off-diagonal $ji$ and $ij$-elements,
respectively, and
$s_{ij}\equiv\sin\theta_{ij}$, $c_{ij}\equiv\cos\theta_{ij}$.
Without loss of generality, we can adopt the convention 
$0\le\theta_{ij}\le \pi/2$, $0\le\delta<2\pi$ \cite{Hagiwara:1998hv}.
Of the six parameters in this expression and the three neutrino masses, 
which add up to a total of nine parameters,
neutrino $\rightarrow$ neutrino oscillations are only sensitive to six:
\begin{itemize}
\item the three mixing angles: $\theta_{12}$, $\theta_{23}$, $\theta_{13}$,
\item two mass-squared differences: $\delta m^2_{21}$, $\delta m^2_{31}$,
where $\delta m^2_{ij} = m^2_i - m^2_j$, and
\item the CP-violating phase: $\delta$.
\end{itemize}
The Majorana phases, $\alpha_{21}$ and $\alpha_{31}$, only appear
in lepton-number violating processes such as neutrinoless double beta decay, 
and cannot be determined via neutrino$\rightarrow$neutrino oscillations.
The absolute scale of the neutrino masses also remain undetermined since
neutrino oscillation is an interference effect.

\subsection{Neutrino Oscillation}
\label{subsec:oscillation}

If a neutrino of flavor $\alpha$ is created at $x=0$ with energy $E$, then
the state of the neutrino at $x=0$ is
\begin{equation}
\ket{\nu_{\alpha,0}(x=0)} 
\;=\; \ket{\nu_\alpha}
\;=\; \sum_{j=1}^3 (V_{\mathrm{PMNS}})^*_{\alpha j} \ket{\nu_j}\;.
\end{equation}
At $x=L$, the same state is
\begin{equation}
\ket{\nu_{\alpha,0}(x=L)}
\;=\; \sum_{j=1}^3 e^{i p_j L}\, (V_{\mathrm{PMNS}})^*_{\alpha j} \ket{\nu_j} 
\;=\; e^{ip_1 L}\sum_{j=1}^3 e^{i(p_j-p_1)L}\, (V_{\mathrm{PMNS}})^*_{\alpha j} \ket{\nu_j} 
\;.
\end{equation}
Assing $m_j\ll E$ we can approximate
\begin{equation}
p_j \;=\; \sqrt{E^2-m_j^2} \;=\; E - \frac{m_j^2}{2E} + \cdots
\end{equation}
so that
\begin{equation}
p_j - p_1 \;\approx\; -\dfrac{\delta m^2_{j1}}{2E}\;,\qquad
\delta m^2_{j1}\;=\; m_j^2 - m_1^2\;,
\end{equation}
and we find
\begin{equation}
\ket{\nu_{\alpha,0}(x=L)}
\;=\; e^{i p_1 L}\sum_{j=1}^3 
\exp\!\left(-i\dfrac{\delta m_{j1}^2}{2E}L\right)\,
(V_{\mathrm{PMNS}})^*_{\alpha j}\ket{\nu_j}\;.
\end{equation}
Therefore, the amplitude of observing the neutrino of flavor $\beta$ at
$x=L$ is given by (dropping the irrelevant overall phase)
\begin{eqnarray}
\mathcal{A}_{\beta\alpha}
& = &
\braket{\nu_\beta}{\nu_{\alpha,0}(x=L)} \phantom{\Bigg|} \cr
& = & 
\Biggl[\sum_{k=1}^3\bra{\nu_k}(V_{\mathrm{PMNS}})_{\beta k} 
\Biggr]
\Biggl[\sum_{j=1}^3 
\exp\!\left(-i\dfrac{\delta m_{j1}^2}{2E}L\right)\,
(V_{\mathrm{PMNS}})^*_{\alpha j}\ket{\nu_j}
\Biggr] \cr
& = & \sum_{j=1}^3
(V_{\mathrm{PMNS}})_{\beta j}\;\exp\!\left(-i\dfrac{\delta m_{j1}^2}{2E}L\right)\,
(V_{\mathrm{PMNS}})^*_{\alpha j}
\cr
& = & \sum_{j=1}^3
\,U_{\beta j}\;\exp\!\left(-i\dfrac{\delta m_{j1}^2}{2E}L\right)\,
U^*_{\alpha j}
\cr
& = &
\left[\,
U\,\exp\!\left(-i\dfrac{\delta M^2}{2E}L\right)
U^\dagger
\right]_{\beta\alpha}
\cr
& = &
\left[\,
\exp\!\left(-i\dfrac{H_0}{2E}L\right)
\right]_{\beta\alpha}
\;,
\end{eqnarray}
where
\begin{equation}
\delta M^2 
\;=\;
\left[\begin{array}{ccc}
0 & 0 & 0 \\
0 & \delta m^2_{21} & 0 \\
0 & 0 & \delta m^2_{31}
\end{array}\right]
\;,
\end{equation}
and
\begin{equation}
H_0 \;=\; U\,\delta M^2\,U^\dagger \;.
\end{equation}
%
Thus,
the probability of oscillation from $\ket{\nu_\alpha}$ to
$\ket{\nu_\beta}$ with neutrino energy $E$ and baseline $L$ is given by
\begin{eqnarray}
P(\nu_\alpha\rightarrow\nu_\beta)
& = &
\bigl|\,\mathcal{A}_{\beta\alpha}\,\bigr|^2
\cr
& = &
\left| \sum_{j=1}^3 U_{\beta j}^{\phantom{*}} \,
       \exp\!\left( -i\dfrac{\delta m_{j1}^2}{2E} L\right) U^*_{\alpha j}
\right|^2 \cr
& = & \delta_{\alpha\beta}
      -4\sum_{i>j}
       \Re\left(U^*_{\alpha i}U_{\beta i}^{\phantom{*}}U_{\alpha j}^{\phantom{*}}U^*_{\beta j}\right)\sin^2\frac{\Delta_{ij}}{2} 
      +2\sum_{i>j}
       \Im\left(U^*_{\alpha i}U_{\beta i}^{\phantom{*}}U_{\alpha j}^{\phantom{*}}U^*_{\beta j}\right)\sin\Delta_{ij} \;, \cr
& & \label{Palphatobeta}
\end{eqnarray}
where\footnote{%
Note that our notation differs from that of Cervera et al. in Ref.~\cite{Cervera:2000kp}.
There, the symbol $\Delta_{ij}$ is defined without the factor of $L$, that is,
$\Delta_{ij}=\delta m_{ij}^2/2E$.
It also differs from that used by Freund in Ref.~\cite{Freund:2001pn}
where $\Delta = \delta m_{31}^2$, and $\hat{\Delta}=\delta m^2_{31}L/4E$.
Huber and Winter in Ref.~\cite{Huber:2003ak} define $\Delta = \delta m_{31}^2 L/4E$,
which is also used in Ref.~\cite{Nakamura:2012}.
So care is necessary when comparing formulae.
}
\begin{equation}
\Delta_{ij} 
\;\equiv\; \dfrac{\delta m_{ij}^2}{2E} L 
\;=\; 2.534\;
\biggl(\dfrac{\delta m_{ij}^2}{\mathrm{eV}^2}\biggr)
\biggl(\dfrac{\mathrm{GeV}}{E}\biggr)
\biggl(\dfrac{L}{\mathrm{km}}\biggr)
\;,\qquad
\delta m_{ij}^2
\,\equiv\, m_i^2-m_j^2\;.
\end{equation}
%
%
Since
\begin{equation}
\Delta_{32} = \Delta_{31} - \Delta_{21}\;,
\end{equation}
only two of the three $\Delta_{ij}$'s in Eq.~(\ref{Palphatobeta}) are independent.  
Eliminating $\Delta_{32}$ from Eq.~(\ref{Palphatobeta})
for the $\alpha=\beta$ case yields
\begin{eqnarray}
P(\nu_\alpha\rightarrow\nu_\alpha)
& = & 1 - 4\, |U_{\alpha 2}|^2 \left( 1 - |U_{\alpha 2}|^2 \right)
            \sin^2\frac{\Delta_{21}}{2}
        - 4\, |U_{\alpha 3}|^2 \left( 1 - |U_{\alpha 3}|^2 \right)
            \sin^2\frac{\Delta_{31}}{2} \cr
& &  \phantom{1}
        + 2\, |U_{\alpha 2}|^2 |U_{\alpha 3}|^2
          \left( 4\sin^2\frac{\Delta_{21}}{2}\sin^2\frac{\Delta_{31}}{2}
                + \sin\Delta_{21}\sin\Delta_{31}
          \right) \;,
\label{Palphatoalpha}
\end{eqnarray}
and for the $\alpha\neq\beta$ case we have
\begin{eqnarray}
P(\nu_\alpha \rightarrow \nu_\beta)
& = & 4\, |U_{\alpha 2}|^2 |U_{\beta 2}|^2 \sin^2\frac{\Delta_{21}}{2}
     +4\, |U_{\alpha 3}|^2 |U_{\beta 3}|^2 \sin^2\frac{\Delta_{31}}{2} \cr
&   & +2\;\Re\left( U^*_{\alpha 3}U_{\beta 3}U_{\alpha 2}U^*_{\beta 2}\right)
      \left(4\sin^2\frac{\Delta_{21}}{2}\sin^2\frac{\Delta_{31}}{2}
           +\sin\Delta_{21}\sin\Delta_{31}
      \right) \cr 
&   & +4\,J_{(\alpha,\beta)}
      \left( \sin^2\frac{\Delta_{21}}{2}\sin\Delta_{31}
            -\sin^2\frac{\Delta_{31}}{2}\sin\Delta_{21}
      \right) \;,    
\label{Palphatonotalpha}       
\end{eqnarray}
where $J_{(\alpha,\beta)}$ is the Jarlskog invariant \cite{Jarlskog:1985ht}:
\begin{eqnarray}
J_{(\alpha,\beta)}
& = & +\Im(U^*_{\alpha 1}U_{\beta 1}U_{\alpha 2}U^*_{\beta 2})
\;=\; +\Im(U^*_{\alpha 2}U_{\beta 2}U_{\alpha 3}U^*_{\beta 3})
\;=\; +\Im(U^*_{\alpha 3}U_{\beta 3}U_{\alpha 1}U^*_{\beta 1}) \cr
& = & -\Im(U^*_{\alpha 2}U_{\beta 2}U_{\alpha 1}U^*_{\beta 1})
\;=\; -\Im(U^*_{\alpha 1}U_{\beta 1}U_{\alpha 3}U^*_{\beta 3})
\;=\; -\Im(U^*_{\alpha 3}U_{\beta 3}U_{\alpha 2}U^*_{\beta 2}) \cr
& = & -J_{(\beta,\alpha)}\;.
\end{eqnarray}
In the parametrization given in Eq.~(\ref{UPparam}), we have
\begin{equation}
J_{(\mu,e)} \;=\; -J_{(e,\mu)} \;=\;
J_{(e,\tau)} \;=\; -J_{(\tau,e)} \;=\;
J_{(\tau,\mu)} \;=\; -J_{(\mu,\tau)} \;=\; \hat{J}\sin\delta\;, 
\end{equation}
with
\begin{equation}
\hat{J} \;=\; 
s_{12}^{\phantom{2}}
c_{12}^{\phantom{2}}
s_{13}^{\phantom{2}}
c_{13}^2 
s_{23}^{\phantom{2}}
c_{23}^{\phantom{2}}\;.
\end{equation}
The oscillation probabilities for the anti-neutrinos are obtained
by replacing $U_{\alpha i}$ with its complex conjugate, which only amounts to
flipping the sign of $\delta$ in the parametrization of Eq.~(\ref{UPparam}).
It is clear from Eq.~(\ref{Palphatoalpha}) that
$P(\overline{\nu}_\alpha\rightarrow\overline{\nu}_\alpha) = P(\nu_\alpha\rightarrow\nu_\alpha)$,
which is to be expected from the CPT theorem.
For flavor changing oscillations, only the Jarskog term in
Eq.~(\ref{Palphatonotalpha}) changes sign.

\subsection{Matter Effects}
\label{subsec:matter_effect}


If the matter density along the baseline is constant, matter effects 
on neutrino oscillations can be taken into account by 
replacing the PMNS matrix elements and mass-squared 
differences with their ``effective'' values in matter:
\begin{equation}
\Delta_{ij}\rightarrow\tilde{\Delta}_{ij}\;,\quad
U_{\alpha i}\rightarrow \tilde{U}_{\alpha i}\;,
\end{equation}
where $\tilde{U}$ is the unitary matrix that diagonalizes the modified
Hamiltonian,
\begin{equation}
H_a \;=\; 
\tilde{U}
\left[ \begin{array}{ccc} \lambda_1 & 0 & 0 \\
                          0 & \lambda_2 & 0 \\
                          0 & 0 & \lambda_3
       \end{array}
\right]
\tilde{U}^\dagger
\;=\; 
\underbrace{
U
\underbrace{
\left[ \begin{array}{ccc} 0 & 0 & 0 \\
                          0 & \delta m^2_{21} & 0 \\
                          0 & 0 & \delta m^2_{31}
       \end{array}
\right]
}_{\displaystyle =\delta M^2}
U^\dagger}_{\displaystyle =H_0}
+
\left[ \begin{array}{ccc} a & 0 & 0 \\
                          0 & 0 & 0 \\
                          0 & 0 & 0 
       \end{array}
\right] \;,
\label{HinMatter}
\end{equation}
and
\begin{equation}
\tilde{\Delta}_{ij}
\;=\; \dfrac{\delta\lambda_{ij}}{2E}\,L\;,\qquad
\delta\lambda_{ij} = \lambda_i - \lambda_j\;.
\end{equation}
The factor $a$ is due to the interaction of the $\ket{\nu_e}$ component
of the neutrinos with the electrons in matter via $W$-exchange:
\begin{equation}
a \;=\; 2\sqrt{2}\,G_F N_e E \;.
\end{equation}
Assuming $N_e=N_p\approx N_n$ in Earth matter, 
$N_e$ for mass density per unit volume of $\rho$ can be expressed using 
Avogadro's number $N_A=6.02214129\times 10^{23}\,\mathrm{mol}^{-1}$
as
\begin{equation}
N_e \;=\; N_p
\;\approx\; \rho\,N_A/2
\;=\; \left(3.011\times 10^{23}\,\mathrm{/cm^3}\right)
\times\left(\dfrac{\rho}{\mathrm{g/cm^3}}\right)
\;.
\end{equation}
Thus, putting back powers of $\hbar c$ to convert from natural to conventional units, we find
\begin{eqnarray}
a 
& = & 2\sqrt{2}\,G_F N_e E \times(\hbar c)^3 
\cr
& = & \left(7.63\times 10^{-5}\,\mathrm{eV}^2\right)
\left(\dfrac{\rho}{\mathrm{g/cm^3}}\right)
\left(\dfrac{E}{\mathrm{GeV}}\right)\;.
\end{eqnarray}
For anti-neutrino beams, $a$ is replaced by $-a$ in Eq.~(\ref{HinMatter}).
Note that $a$ is $E$-dependent, which means that both $\tilde{U}$ and
$\tilde{\Delta}_{ij}$ are also $E$-dependent.
It is also assumed that $E\ll M_W$ since the $W$-exchange interaction is
approximated by a point-like four-fermion interaction in deriving this expression.
%

\section{Jacobi Method}
\label{sec:Jacobi}

\subsection{Setup}

As mentioned in the introduction, it is possible to write down exact analytical
expressions for $\tilde{\Delta}_{ij}$ and $\tilde{U}_{\alpha i}$ \cite{Zaglauer:1988gz}.
However, simpler and more transparent approximate expressions can be obtained
using the Jacobi method as will be shown in the following.

We introduce the matrix
\begin{equation}
\mathcal{Q} \;=\; \mathrm{diag}(1,1,e^{i\delta})\;,
\end{equation}
and start with the partially diagonalized Hamiltonian:
\begin{eqnarray}
H'_a 
& = & \mathcal{Q}^\dagger U^\dagger H_a U\mathcal{Q} \cr
& = & \mathcal{Q}^\dagger \left\{\;
\left[ \begin{array}{ccc} 0 & 0 & 0 \\
                          0 & \delta m^2_{21} & 0 \\
                          0 & 0 & \delta m^2_{31}
       \end{array}
\right] +
U^\dagger
\left[ \begin{array}{ccc} a & 0 & 0 \\
                          0 & 0 & 0 \\
                          0 & 0 & 0 
       \end{array}
\right] 
U \;\right\} \mathcal{Q} \cr
& = & \mathcal{Q}^\dagger
\left[ \begin{array}{ccc} 0 & 0 & 0 \\
                          0 & \delta m^2_{21} & 0 \\
                          0 & 0 & \delta m^2_{31}
       \end{array}
\right] \mathcal{Q}
+ a \,\mathcal{Q}^\dagger
\left[ \begin{array}{ccc} U^*_{e1}U_{e1} & U^*_{e1}U_{e2} & U^*_{e1}U_{e3} \\
                          U^*_{e2}U_{e1} & U^*_{e2}U_{e2} & U^*_{e2}U_{e3} \\
                          U^*_{e3}U_{e1} & U^*_{e3}U_{e2} & U^*_{e3}U_{e3} 
       \end{array}
\right] \mathcal{Q} \cr
& = &
\left[ \begin{array}{ccc} 0 & 0 & 0 \\
                          0 & \delta m^2_{21} & 0 \\
                          0 & 0 & \delta m^2_{31}
       \end{array}
\right] + a 
\left[ \begin{array}{ccc} 
        c_{12}^2 c_{13}^2    & c_{12}s_{12}c_{13}^2 & c_{12}c_{13}s_{13}  \\
        c_{12}s_{12}c_{13}^2 & s_{12}^2c_{13}^2     & s_{12}c_{13}s_{13}  \\
        c_{12}c_{13}s_{13}   & s_{12}c_{13}s_{13}   & s_{13}^2 
       \end{array}
\right] \cr
& = & 
\left[ 
\begin{array}{ccc} 
a c_{12}^2 c_{13}^2    & a c_{12}s_{12}c_{13}^2 & a c_{12}c_{13}s_{13} \\
a c_{12}s_{12}c_{13}^2 & a s_{12}^2c_{13}^2 + \delta m^2_{21} & a s_{12}c_{13}s_{13} \\
a c_{12}c_{13}s_{13}   & a s_{12}c_{13}s_{13}   & a s_{13}^2 + \delta m^2_{31} 
\end{array}
\right] 
\;.
\label{Hprime}
\end{eqnarray}
%
The matrix $\mathcal{Q}$ serves to rid $H'_a$ of any reference to the
CP violating phase $\delta$.
The strategy we used in our previous papers \cite{Honda:2006hp,Honda:2006gv} was to 
approximately diagonalize $H'_a$ through the Jacobi method
using
\begin{equation}
\varepsilon
\;=\; \sqrt{\dfrac{\delta m^2_{21}}{|\delta m^2_{31}|}} 
\;\approx\; 0.17\;,
\label{varepsilondef}
\end{equation}
as the parameter to keep track of the sizes of the off-diagonal elements.
We argued that approximate diagonalization was achieved when the off-diagonal
elements were of order $O(\varepsilon^2 s_{13}|\delta m^2_{31}|)$.

Note that our $\varepsilon$ differs from Asano and Minakata's $\epsilon$ 
in Ref.~\cite{Asano:2011nj} where
\begin{equation}
\epsilon
\;=\;\dfrac{\delta m^2_{21}}{|\delta m^2_{31}|}
\;\approx\; 0.03\;.
\end{equation}
That is, $\epsilon=\varepsilon^2$.
So care is necessary when comparing formulae.

\subsection{Diagonalization of a 2 $\times$ 2 Matrix}

Recall that for $2\times 2$ real symmetric matrices, such as
\begin{equation}
M = \left[ \begin{array}{cc} \alpha & \beta \\ \beta & \gamma \end{array}\right]\;,\qquad
\alpha, \beta, \gamma \in\mathbb{R}\;,
\end{equation}
diagonalization is trivial. Just define
\begin{equation}
R = \left[ \begin{array}{rr} c_\omega & s_\omega \\
                            -s_\omega & c_\omega 
           \end{array}
    \right]\;,\quad
\mbox{where}\qquad
c_\omega = \cos\omega\;,\qquad
s_\omega = \sin\omega\;,\qquad
\tan 2\omega \equiv \dfrac{2\beta}{\gamma-\alpha}\;,
\end{equation}
and we obtain
\begin{equation}
R^\dagger M R
= \left[ \begin{array}{cc} \Lambda_1 & 0 \\ 0 & \Lambda_2 \end{array}\right]\;,
\end{equation} 
with
\begin{eqnarray}
\Lambda_1 
& = & \dfrac{\alpha c^2_\omega - \gamma s^2_\omega}{c^2_\omega-s^2_\omega}
\;=\; \dfrac{(\alpha+\gamma)\mp\sqrt{(\alpha-\gamma)^2+4\beta^2}}{2}\;,\cr
\Lambda_2 
& = & \dfrac{\gamma c^2_\omega - \alpha s^2_\omega}{c^2_\omega-s^2_\omega}
\;=\; \dfrac{(\alpha+\gamma)\pm\sqrt{(\alpha-\gamma)^2+4\beta^2}}{2}\;,
\end{eqnarray}
where the upper and lower signs are for the cases $\alpha<\gamma$ and $\alpha>\gamma$, respectively.
The Jacobi method \cite{Jacobi:1846} entails iteratively diagonalizing $2\times 2$ submatrices
of a larger matrix in the order that requires the largest rotation angle at each step.
In the limit of infinite iterations of this procedure, the matrix will
converge to a diagonal matrix.

In the case of $H'_a$ given in Eq.~(\ref{Hprime}), 
at most two iterations are sufficient to
achieve approximate diagonalization,
neglecting off-diagonal elements of order 
$O(\varepsilon^2 s_{13}|\delta m^2_{31}|)$,
regardless of the size of $a$.
We demonstrate this in this appendix.

\subsection{Neutrino Case}
\subsubsection{Mixing Angles and Mass-squared Differences}

Let us first evaluate the sizes of the sines and cosines of the
three vacuum mixing angles in comparison to 
the parameter $\varepsilon$ defined in Eq.~(\ref{varepsilondef}).
The current best fit values for the mass-squared differences and
mixing angles are listed in Table~\ref{tab:bench}.
The sines and cosines of the central values of the mixing angles are
\begin{equation}
\begin{array}{ll}
s_{23} \;=\; 0.71 \;,\quad &
c_{23} \;=\; 0.71 \;,\\ 
s_{12} \;=\; 0.55 \;,\quad &
c_{12} \;=\; 0.84 \;,\\ 
s_{13} \;=\; 0.15 \;,\quad &
c_{13} \;=\; 0.99 \;.\\
\end{array}
\label{scorders}
\end{equation}
%
%
Therefore, $s_{13}$ is $O(\varepsilon)$ while all other
sines and cosines are $O(1)$.

\subsubsection{First rotation}

The effective hamiltonian we need to diagonalize is
\begin{eqnarray}
H'_a & = &
\left[
	\begin{array}{ccc}
	a c_{12}^2 c_{13}^2 & a c_{12} s_{12} c_{13}^2 & a c_{12} c_{13} s_{13} \\
	a c_{12} s_{12} c_{13}^2 & a s_{12}^2 c_{13}^2 + \delta m_{21}^2 & a s_{12} c_{13} s_{13} \\
	a c_{12} c_{13} s_{13} & a s_{12} c_{13} s_{13} & a s_{13}^2 + \delta m_{31}^2
	\end{array}
\right]
\cr
& = &
\left[
	\begin{array}{ccc}
	a\mathcal{O}(1) & a\mathcal{O}(1) & a\mathcal{O}(\varepsilon) \\
	a\mathcal{O}(1) & a\mathcal{O}(1) + \delta m_{21}^2 & a\mathcal{O}(\varepsilon) \\
	a\mathcal{O}(\varepsilon) & a\mathcal{O}(\varepsilon) & a\mathcal{O}(\varepsilon^2) + \delta m_{31}^2
	\end{array}
\right] 
\;.
\end{eqnarray}
Of the off-diagonal elements, 
the 1-2 element is the largest regardless of the size of $a$. 
Therefore, our first step is to 
diagonalize the 1-2 submatrix.

\begin{figure}[t]
\subfigure[]{\includegraphics[height=5.1cm]{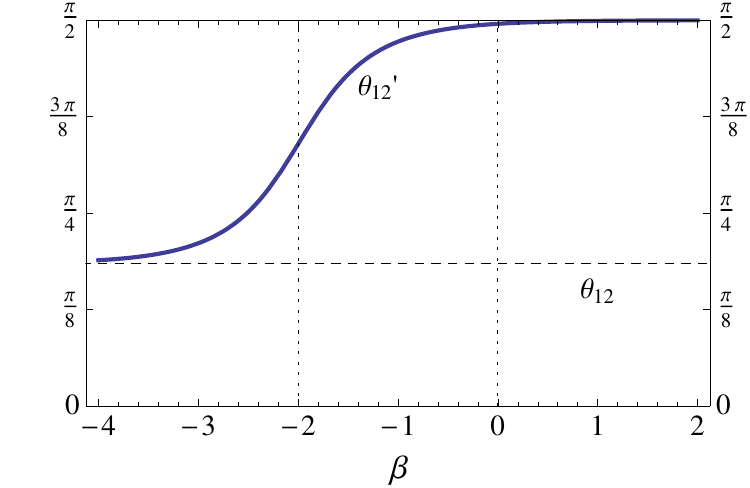}}
\subfigure[]{\includegraphics[height=5cm]{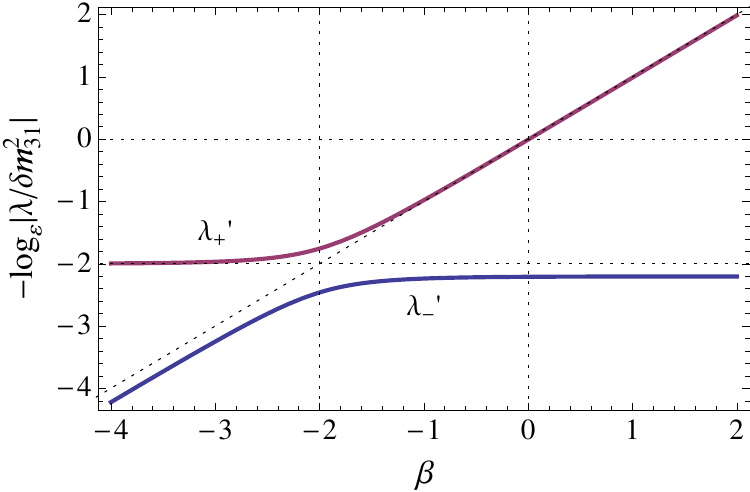}}
\caption{(a) The dependence of $\theta'_{12}$ on 
$\beta=-\log_{\,\varepsilon}\left(a/|\delta m^2_{31}|\right)$.
(b) The $\beta$-dependence of $\lambda'_{\pm}$.
}
\label{theta12primeplot}
\end{figure}
%
\begin{figure}[b]
\subfigure[]{\includegraphics[height=5cm]{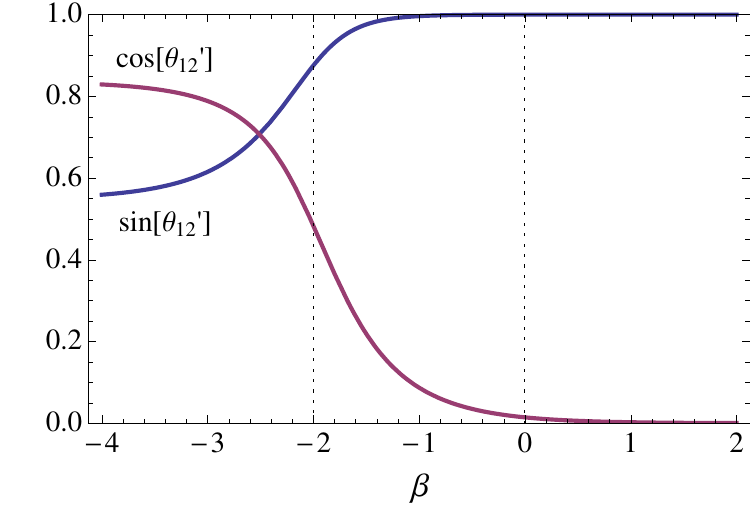}}
\subfigure[]{\includegraphics[height=5cm]{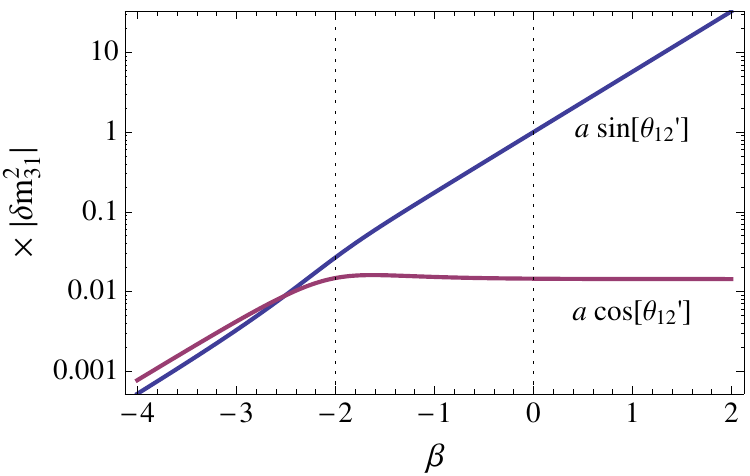}}
\caption{(a) The dependence of $s'_{12}=\sin\theta'_{12}$ and
$c'_{12}=\cos\theta'_{12}$ on 
$\beta=-\log_{\,\varepsilon}\left(a/|\delta m^2_{31}|\right)$.
(b) The dependence of $a s'_{12}$ and $a c'_{12}$ on $\beta$.
The values are given in units of $|\delta m^2_{31}|$.
The asymptotic value of $a c'_{12}$ is $\delta m^2_{21}s_{12}c_{12}/c_{13}^2
\approx 0.014\;|\delta m^2_{31}| = O(\varepsilon^2|\delta m^2_{31}|)$.
}
\label{s12primec12primeplot}
\end{figure}

Define
\begin{equation}
V \;\equiv\; 
\left[
\begin{array}{ccc}
c_{\varphi} & s_{\varphi} & 0  \\
-s_{\varphi} & c_{\varphi} & 0 \\
0 & 0 & 1
\end{array}
\right],
\end{equation}
where
\begin{equation}
c_{\varphi} \;=\; \cos \varphi \,, \quad
s_{\varphi} \;=\; \sin \varphi \,, \quad
\tan 2\varphi \;\equiv\; 
	\dfrac{a c_{13}^2 \sin 2\theta_{12} }{ \delta m_{21}^2 - a c_{13}^2 \cos 2 \theta_{12} } \,, \quad
\left(0 \le \varphi \le \frac{\pi}{2} \right)
\;.
\end{equation}
Using $V$, we find
\begin{equation}
H''_a \;\equiv\; V^\dagger H'_a V \;=\; 
\left[
	\begin{array}{ccc}
	\lambda'_- & 0 & a c_{12}' c_{13} s_{13} \\
	0 & \lambda_+' & a s_{12}' c_{13} s_{13} \\
	a c_{12}' c_{13} s_{13} & a s_{12}' c_{13} s_{13} & a s_{13}^2 + \delta m_{31}^2 
	\end{array}
\right] \, ,
\label{Hdoubleprime}
\end{equation}
where
\begin{equation}
c_{12}' \;=\; \cos \theta_{12}' \;, \quad
s_{12}' \;=\; \sin \theta_{12}' \;, \quad
\theta_{12}' \;\equiv\; \theta_{12} + \varphi \;,
\end{equation}
and
%
%
%
\begin{equation}
\lambda_{\pm}' \;\equiv\; 
\dfrac{ ( \delta m_{21}^2 + a c_{13}^2 ) \pm 
		\sqrt{ ( \delta m_{21}^2 - a c_{13}^2 )^2 + 4a c_{13}^2 s_{12}^2 \delta m_{21}^2 } }
			 { 2 } \;.
\end{equation}
The angle $\theta'_{12}=\theta_{12}+\varphi$ can be calculated directly without
calculating $\varphi$ via
\begin{equation}
\tan 2\theta'_{12} \;=\; 
\dfrac{\delta m^2_{21} \sin 2\theta_{12}}{\delta m^2_{21}\cos 2\theta_{12} - ac^2_{13}}
\;,\qquad
\left(\theta_{12}\le\theta'_{12}\le\dfrac{\pi}{2}\right)\;.
\end{equation}
The dependences of $\theta'_{12}$ and $\lambda'_{\pm}$ 
on $\beta=-\log_{\varepsilon}\left(a/|\delta m^2_{31}|\right)$ 
are plotted in Fig.~\ref{theta12primeplot}.
Note that $\theta'_{12}$ increases monotonically from $\theta_{12}$ to $\pi/2$ with increasing $a$.
The $\beta$-dependence of $s'_{12}=\sin\theta'_{12}$ and $c'_{12}=\cos\theta'_{12}$
are shown in Fig.~\ref{s12primec12primeplot}(a).
For $a\gg \delta m^2_{21}$, $s'_{12}$ and $c'_{12}$ behave as
\begin{eqnarray}
s'_{12} & = & 1
- \dfrac{s_{12}^2 c_{12}^2}{2}\left(\dfrac{\delta m^2_{21}}{a c_{13}^2}\right)^2
+ \cdots
\;,\cr
c'_{12} & = & 
s_{12}c_{12}\left(\dfrac{\delta m^2_{21}}{a c_{13}^2}\right)
+ s_{12}c_{12}(c_{12}^2-s_{12}^2)\left(\dfrac{\delta m^2_{21}}{a c_{13}^2}\right)^2
+ \cdots
\;.
\label{s12primec12prime}
\end{eqnarray}
Therefore, for $a\gg \delta m^2_{21}$ we have
$a s'_{12}\approx a $ while 
$a c'_{12}\approx \delta m^2_{21}s_{12}c_{12}/c_{13}^2
\;=\; \varepsilon^2|\delta m^2_{31}|s_{12}c_{12}/c_{13}^2
\approx 0.014\;|\delta m^2_{31}| = O(\varepsilon^2|\delta m^2_{31}|)$.
This behavior is shown in Fig.~\ref{s12primec12primeplot}(b).
Note that $a c'_{12}$ never grows larger than $O(\varepsilon^2|\delta m^2_{31}|)$
for any $a$.

The values of $\lambda'_\pm$ away from the level crossing point
$a\sim \delta m^2_{21}$ for the $a\ll\delta m^2_{21}$ case are given by
\begin{eqnarray}
\lambda'_-
& = & a c^2_{13}c^2_{12}
\left[ 
1
- s^2_{12}\left(\dfrac{a c^2_{13}}{\delta m^2_{21}}\right)
- s^2_{12}(c^2_{12}-s^2_{12})\left(\dfrac{a c^2_{13}}{\delta m^2_{21}}\right)^2
+ \cdots
\right]
\;,\cr
\lambda'_+
& = & \delta m^2_{21}
\left[
1 
+ s^2_{12}\left(\dfrac{a c^2_{13}}{\delta m^2_{21}}\right)
+ s^2_{12}c^2_{12}\left(\dfrac{a c^2_{13}}{\delta m^2_{21}}\right)^2
+ s^2_{12}c^2_{12}(c^2_{12}-s^2_{12})\left(\dfrac{a c^2_{13}}{\delta m^2_{21}}\right)^3
+ \cdots
\right]
\;,
\cr
& & 
\label{lambdaprimeexpand1}
\end{eqnarray}
and those for the $a\gg\delta m^2_{21}$ case by
\begin{eqnarray}
\lambda'_-
& = & \delta m^2_{21} c^2_{12}
\left[ 
1
- s^2_{12}\left(\dfrac{\delta m^2_{21}}{a c^2_{13}}\right)
- s^2_{12}(c^2_{12}-s^2_{12})\left(\dfrac{\delta m^2_{21}}{a c^2_{13}}\right)^2
+ \cdots
\right]
\;,\cr
\lambda'_+
& = & a c^2_{13}
\left[
1 
+ s^2_{12}\left(\dfrac{\delta m^2_{21}}{a c^2_{13}}\right)
+ s^2_{12}c^2_{12}\left(\dfrac{\delta m^2_{21}}{a c^2_{13}}\right)^2
+ s^2_{12}c^2_{12}(c^2_{12}-s^2_{12})\left(\dfrac{\delta m^2_{21}}{a c^2_{13}}\right)^3
+ \cdots
\right]
\;.
\cr
& &
\label{lambdaprimeexpand2} 
\end{eqnarray}
We will use this expansion for $\lambda'_+$ later.
Thus, the asymptotic values of $\lambda'_\pm$ are
$\lambda'_-\rightarrow ac^2_{13}c^2_{12}$,
$\lambda'_+\rightarrow \delta m^2_{21}$ in the $a\rightarrow 0$ limit,
and 
$\lambda'_-\rightarrow \delta m^2_{21}c^2_{12}$,
$\lambda'_+\rightarrow ac^2_{13}$ in the $a\rightarrow\infty$ limit.

\subsubsection{Second rotation}

The effective hamiltonian after the first rotation was given 
by Eq.~(\ref{Hdoubleprime}).
When $a< \delta m^2_{21}$, both 
non-zero off-diagonal elements are of order
$O(\varepsilon a) < O(\varepsilon^3|\delta m^2_{31}|)$, since
$s'_{12}$ and $c'_{12}$ are both $O(1)$ in that range
as can be discerned from Fig.~\ref{s12primec12primeplot}(a).
However, as $a$ increases beyond $\delta m^2_{12}$ and
$\theta_{12}'$ approaches $\pi/2$, we have
$a s'_{12}\rightarrow a$, $a c'_{12}\rightarrow O(\varepsilon^2|\delta m^2_{31}|)$ and
the 2-3 element becomes the larger of the two. 
Therefore, a 2-3 rotation is needed next.

\begin{figure}[t]
\subfigure[]{\includegraphics[height=5.1cm]{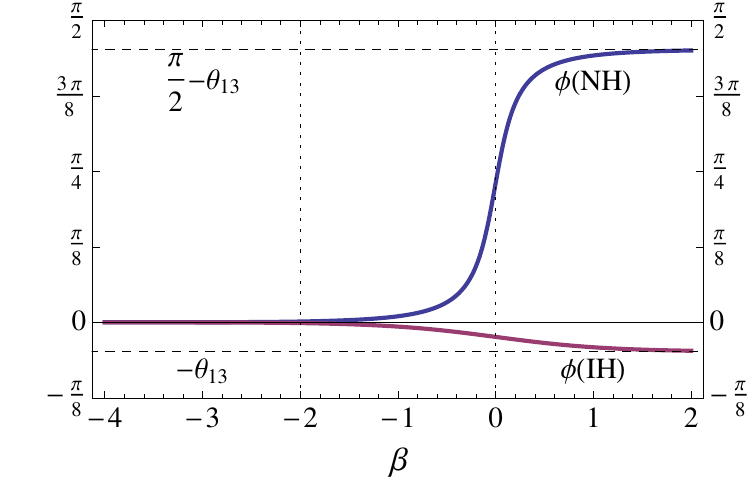}}
\subfigure[]{\includegraphics[height=5cm]{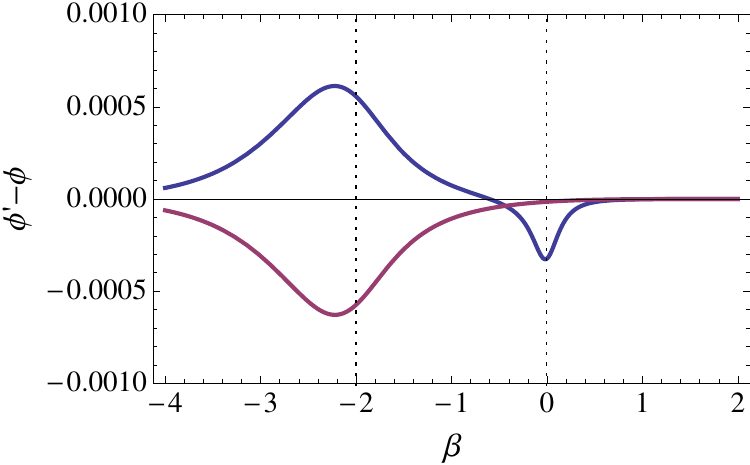}}
\caption{(a) The $\beta$-dependence of $\phi$ for the normal and inverted hierarchies.
(b) The $\beta$-dependence of the difference $\phi'-\phi$.
}
\label{phiphiprimecomparison}
\end{figure}

We define
\begin{equation}
W \;\equiv\; 
\left[
	\begin{array}{ccc}
	1 & 0 & 0 \\
	0 & c_{\phi} & s_{\phi} \\
	0 & -s_{\phi} & c_{\phi}
	\end{array}
\right] \;,
\end{equation}
where
\begin{equation}
c_{\phi} \;=\; \cos \phi \;,\quad
s_{\phi} \;=\; \sin \phi \;,\quad
\tan 2\phi \;\equiv\; 
\dfrac{ a s_{12}' \sin 2\theta_{13} }
      { \delta m_{31}^2 + a s_{13}^2 - \lambda'_+ } \;.
\label{phidef}
\end{equation}
The angle $\phi$ is in the first quadrant when $\delta m_{31}^2 > 0$, 
and in the fourth quadrant when $\delta m_{31}^2 < 0$. Then,
\begin{equation}
H'''_a 
\;\equiv\; W^{\dagger} H''_a W 
\;=\; 
\left[
	\begin{array}{ccc}
	\lambda_{-}' & -a c_{12}' c_{13} s_{13} s_{\phi} & a c_{12}' c_{13} s_{13} c_{\phi} \\
	-a c_{12}' c_{13} s_{13} s_{\phi} & \lambda_{\mp}'' & 0 \\
	a c_{12}' c_{13} s_{13} c_{\phi} & 0 & \lambda_{\pm}''
	\end{array}
\right] \;,
\label{Htripleprime}
\end{equation}
where the upper(lower) sign corresponds to the normal(inverted) hierarchy case
with 
%
%
%
\begin{equation}
\lambda''_{\pm} \;\equiv\;
   \dfrac{ \bigl[\; \lambda'_{+} + (\delta m^2_{31}+a s_{13}^2) \,\bigr]
\pm \sqrt{ \bigl[\; \lambda'_{+} - (\delta m^2_{31}+a s_{13}^2) \,\bigr]^2 
         + 4 a^2 {s^{\prime\;2}_{12}}\,c_{13}^2\, s_{13}^2 }
         }
         { 2 } \;.
\label{lambdadoubleprimeplusminusdef}
\end{equation}
%
%
%
The $\beta$-dependences of $\lambda''_{\pm}$ and $\phi$ are shown 
in Fig.~\ref{lambdadoubleprimeplot} ((a) and (b)), and Fig.~\ref{phiphiprimecomparison}(a),
respectively, for both mass hierarchies.
For the normal hierarchy case, $\delta m^2_{31}>0$,
the values of $\lambda''_{\pm}$ away from the level crossing point $a\sim \delta m^2_{31}$ 
are approximately
\begin{eqnarray}
\lambda''_{+}
& \approx &
\delta m^2_{31} + a s^2_{13}
\;,
\cr
\lambda''_{-}
& \approx &
\lambda'_+
\;,
\end{eqnarray}
when $a\ll\delta m^2_{31}$, and
\begin{eqnarray}
\lambda''_{+}
& \approx &
a + s^2_{13}\delta m^2_{31} + c^2_{13}s^2_{12}\delta m^2_{21}
\;,\cr
\lambda''_{-}
& \approx & 
c^2_{13}\delta m^2_{31} + s^2_{13} s^2_{12} \delta m^2_{21}
\;,
\end{eqnarray}
when $a\gg\delta m^2_{31}$.
For the inverted hierarchy case, $\delta m^2_{31}<0$, where
there is no level crossing, the values of $\lambda''_{\pm}$
are approximately
\begin{equation}
\lambda''_{-} \;\approx\; \delta m^2_{31} \;<\; 0\;,\qquad
\lambda''_{+} \;\approx\; \lambda'_+ \;,
\end{equation}
for all $a$.


At this point, we argue that the angle $\phi$ defined in 
Eq.~(\ref{phidef}) is well approximated by the angle $\phi'$
which we define via
\begin{equation}
\tan 2\phi'
\;\equiv\;
\dfrac{a\sin 2\theta_{13}}
{(\delta m^2_{31}-\delta m^2_{21} s_{12}^2)-a\cos 2\theta_{13}}
\;.
\label{phiprimedef}
\end{equation}
This approximation is obtained by first noting that $\phi$ is significantly different from zero
only when $a\gg \delta m^2_{21}$.
The expansion of $\lambda'_+$ 
in the denominator of the right-hand-side of Eq.~(\ref{phidef}) in powers of 
$\delta m^2_{21}/a$ was given in
Eq.~(\ref{lambdaprimeexpand2}).
Keeping only the first two terms, and
noting also that $s_{12}'\approx 1$ to the same order 
when $a\gg \delta m^2_{21}$ (c.f. Eq.~(\ref{s12primec12prime}))
we obtain Eq.~(\ref{phiprimedef}).
The $\beta$-dependence of the 
difference $\phi'-\phi$ is plotted in Fig.~\ref{phiphiprimecomparison}(b),
and we can see that the disagreement is at most $O(\varepsilon^4)$.
Thus, we replace $\phi$ with $\phi'$ in the following.

Now, the effective Hamiltonian after the second rotation was 
given by Eq.~(\ref{Htripleprime}).
Note that all of the non-zero off-diagonal elements include the factor $a c'_{12}$,
which is never larger than $O(\varepsilon^2|\delta m^2_{31}|)$ regardless of the value of $a$ as discussed above.
They also all include a factor of $s_{13}$, which is $O(\varepsilon)$ as we have seen in Eq.~(\ref{scorders}).
Therefore, all off-diagonal elements of $H_a'''$ are of order 
$O(\varepsilon^2 s_{13}|\delta m^2_{31}|)=O(\varepsilon^3|\delta m^2_{31}|)$
or smaller regardless of the size of $a$.
Note that had the value of $s_{13}$ been smaller, the sizes of the neglected terms would have been
proportionately smaller also. 
We conclude that, at this point, 
off-diagonal elements are negligible and a third rotation is not necessary.

\subsubsection{Absorption of $\phi'$ into $\theta_{13}$}

From the above consideration, we conclude that the matrix which diagonalizes $H_a'$, Eq.~(\ref{Hprime}), is given approximately by $VW$, and 
that the effective neutrino mixing matrix becomes
\begin{equation}
\tilde{U}
\;\approx\;  U\mathcal{Q}VW 
\;=\;
\underbrace{R_{23}(\theta_{23},0)R_{13}(\theta_{13},\delta)R_{12}(\theta_{12},0)}_{\displaystyle U}
\mathcal{Q}
\underbrace{R_{12}(\varphi,0)}_{\displaystyle V}
\underbrace{R_{23}(\phi',0)}_{\displaystyle W} 
\;.
\end{equation}
Using
\begin{eqnarray}
R_{12}(\theta_{12},0)\mathcal{Q} & = & \mathcal{Q}R_{12}(\theta_{12},0) \;,\cr
R_{13}(\theta_{13},\delta)\mathcal{Q} & = & \mathcal{Q}R_{13}(\theta_{13},0) \;,
\end{eqnarray}
we find
\begin{eqnarray}
\tilde{U}
& \approx & 
R_{23}(\theta_{23},0)R_{13}(\theta_{13},\delta)R_{12}(\theta_{12},0)
\mathcal{Q}\,R_{12}(\varphi,0)R_{23}(\phi',0) 
\cr
& = & 
R_{23}(\theta_{23},0)\mathcal{Q}\,
R_{13}(\theta_{13},0)R_{12}(\theta_{12},0)R_{12}(\varphi,0)R_{23}(\phi',0) 
\cr
& = & 
R_{23}(\theta_{23},0)\mathcal{Q}\,
R_{13}(\theta_{13},0)R_{12}(\theta_{12}+\varphi,0)R_{23}(\phi',0) 
\cr
& = & 
R_{23}(\theta_{23},0)\mathcal{Q}\,
R_{13}(\theta_{13},0)R_{12}(\theta'_{12},0)R_{23}(\phi',0) 
\;.
\end{eqnarray}
Here, we argue that
\begin{equation}
R_{12}(\theta'_{12},0)R_{23}(\phi',0) 
\;\approx\;
R_{13}(\phi',0)R_{12}(\theta'_{12},0) 
\;,
\label{Rcommute}
\end{equation}
that is, the 2-3 rotation becomes a 1-3 rotation 
when commuted through $R_{12}(\theta'_{12},0)$.
This is due to the fact that $\phi'$ only becomes non-negligible
when $a\gg \delta m^2_{12}$ where $s'_{12}\approx 1$ and $c'_{12}\approx 0$,
which means
\begin{equation}
R_{12}(\theta'_{12},0)\;\approx\;
\left[\begin{array}{ccc}
 0 & 1 & 0 \\
-1 & 0 & 0 \\  
 0 & 0 & 1 
\end{array}\right]
\;,
\end{equation}
and it is straightforward to see that
\begin{equation}
\left[\begin{array}{ccc}
 0 & 1 & 0 \\
-1 & 0 & 0 \\  
 0 & 0 & 1 
\end{array}\right]
\left[\begin{array}{ccc}
1 & 0 & 0 \\
0 &  c'_{\phi} & s'_{\phi} \\
0 & -s'_{\phi} & c'_{\phi}
\end{array}\right]
\;=\;
\left[\begin{array}{ccc}
 c'_{\phi} & 0 & s'_{\phi} \\
 0 & 1 & 0 \\
-s'_{\phi} & 0 & c'_{\phi}
\end{array}\right]
\left[\begin{array}{ccc}
 0 & 1 & 0 \\
-1 & 0 & 0 \\  
 0 & 0 & 1 
\end{array}\right]
\;,
\end{equation}
where $s'_{\phi}=\sin\phi'$ and $c'_{\phi}=\cos\phi'$.
In the range $a\alt \delta m^2_{21}$, the angle
$\phi'$ is very small and both $R_{23}(\phi',0)$
and $R_{13}(\phi',0)$ are approximately unit matrices and Eq.~(\ref{Rcommute})
is trivially satisfied.
Curiously, this approximation breaks down around $a\sim\delta m^2_{31}$ 
for the normal hierarchy case when $\theta_{13}$ is $O(\varepsilon^2)$ or smaller,
as is discussed in appendix~\ref{sec:Commute}.
However, given that the current experimentally preferred value of $\theta_{13}$ 
is $O(\varepsilon)$, the approximation is valid.
Thus,
\begin{eqnarray}
\tilde{U}
& \approx & 
R_{23}(\theta_{23},0)\mathcal{Q}\,
R_{13}(\theta_{13},0)R_{12}(\theta'_{12},0)R_{23}(\phi',0) 
\cr
& \approx & 
R_{23}(\theta_{23},0)\mathcal{Q}\,
R_{13}(\theta_{13},0)R_{13}(\phi',0)R_{12}(\theta'_{12},0) 
\cr
& = & 
R_{23}(\theta_{23},0)\mathcal{Q}\,
R_{13}(\theta_{13}+\phi',0)R_{12}(\theta'_{12},0) 
\cr
& = & 
R_{23}(\theta_{23},0)\mathcal{Q}\,
R_{13}(\theta'_{13},0)R_{12}(\theta'_{12},0) 
\cr
& = & 
R_{23}(\theta_{23},0)
R_{13}(\theta'_{13},\delta)R_{12}(\theta'_{12},0) 
\mathcal{Q}
\;,
\end{eqnarray}
where we have defined
\begin{equation}
\theta'_{13} \;\equiv\; \theta_{13}+\phi'\;.
\end{equation}
This angle can be calculated directly without calculating $\phi'$ via
\begin{equation}
\tan 2\theta'_{13} \;=\;
\dfrac{(\delta m^2_{31}-\delta m^2_{21} s_{12}^2)\sin 2\theta_{13}}
{(\delta m^2_{31}-\delta m^2_{21} s_{12}^2)\cos 2\theta_{13}-a}
\;.
\end{equation}
The diagonal phase matrix $\mathcal{Q}$ appearing rightmost in the above matrix product
can be absorbed into the redefinition of the major phases and can be dropped.
Thus, we arrive at our final approximation in which the vacuum mixing angles
are replaced by their effective values in matter
\begin{eqnarray}
\theta_{12} & \rightarrow & \theta'_{12} \;=\; \theta_{12}+\varphi\;,\cr
\theta_{13} & \rightarrow & \theta'_{13} \;=\; \theta_{13}+\phi'\;,\cr
\theta_{23} & \rightarrow & \theta_{23}\;,\cr
\delta      & \rightarrow & \delta\;,
\end{eqnarray}
and the eigenvalues of the effective Hamiltonian are given by 
\begin{eqnarray}
\lambda_{1} & \approx & \lambda'_{-}\;,\cr
\lambda_{2} & \approx & \lambda''_{\mp}\;,\cr
\lambda_{3} & \approx & \lambda''_{\pm}\;.
\end{eqnarray}
Note that of the mixing angles, only $\theta_{12}$ and $\theta_{13}$ are shifted.
$\theta_{23}$ and $\delta$ stay at their vacuum values.

\subsection{Anti-Neutrino Case}
\subsubsection{First Rotation}

For the anti-neutrino case, the matter effect parameter $a$ acquires a minus sign.
Thus, the effective hamiltonian to be diagonalized is
\begin{eqnarray}
\overline{H}'_a
& = & 
\left[
\begin{array}{ccc}
-a c_{12}^2 c_{13}^2 
	& -a c_{12} s_{12} c_{13}^2 
	& -a c_{12} c_{13} s_{13} \\
-a c_{12} s_{12} c_{13}^2 
	& -a s_{12}^2 c_{13}^2 + \delta m_{21}^2 
	& - a s_{12} c_{13} s_{13} \\
-a c_{12} c_{13} s_{13} 
	& - a s_{12} c_{13} s_{13} 
	& -a s_{13}^2 + \delta m_{31}^2
\end{array}
\right]
\cr
& = &
\left[
\begin{array}{ccc}
-a \mathcal{O}(1) 
	& -a \mathcal{O}(1) 
	& -a \mathcal{O}(\varepsilon) \\
-a \mathcal{O}(1) 
	& -a \mathcal{O}(1) + \delta m_{21}^2 
	& -a \mathcal{O}(\varepsilon) \\
-a \mathcal{O}(\varepsilon) 
	& -a \mathcal{O}(\varepsilon) 
	& -a \mathcal{O}(\varepsilon^2) + \delta m_{31}^2
\end{array}
\right] 
\;.
\label{Hbarprime}
\end{eqnarray}
The largest off-diagonal element is the 1-2 element. 
Therefore, our first step is to diagonalize the 1-2 submatrix.

\begin{figure}[t]
\subfigure[]{\includegraphics[height=5.1cm]{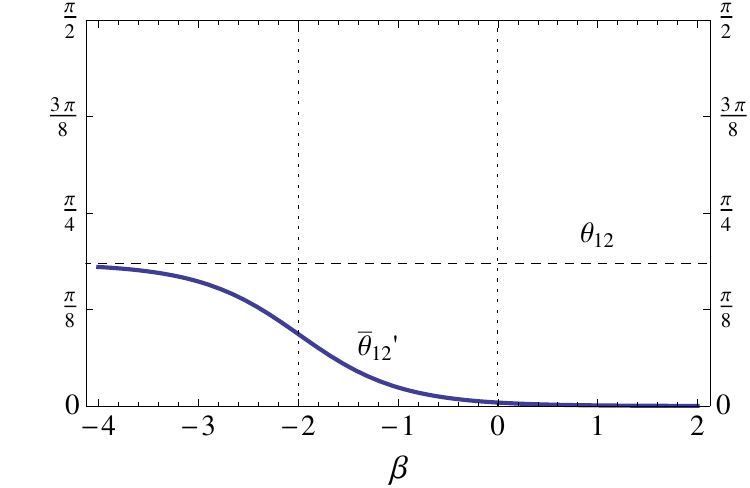}}
\subfigure[]{\includegraphics[height=5cm]{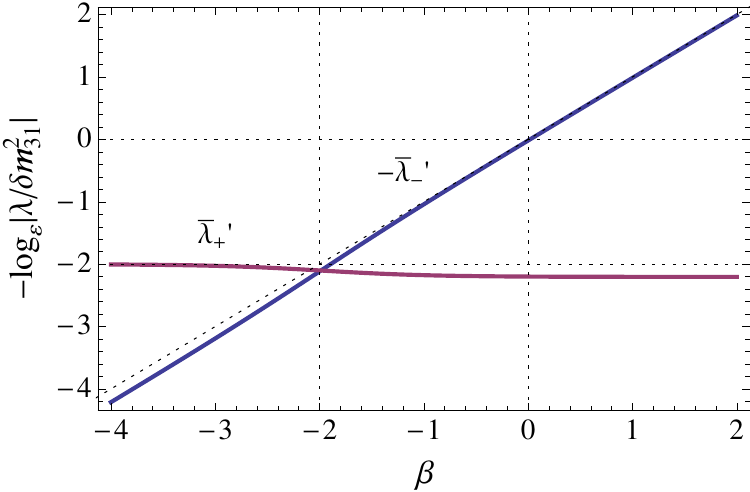}}
\caption{(a) The dependence of $\overline{\theta}'_{12}$ on $\beta=-\log_\varepsilon(a/|\delta m^2_{31}|)$.
(b) The $\beta$-dependence of $\overline{\lambda}'_{\pm}$.
}
\label{theta12primebarplot}
\end{figure}
%

\begin{figure}[t]
\subfigure[]{\includegraphics[height=5cm]{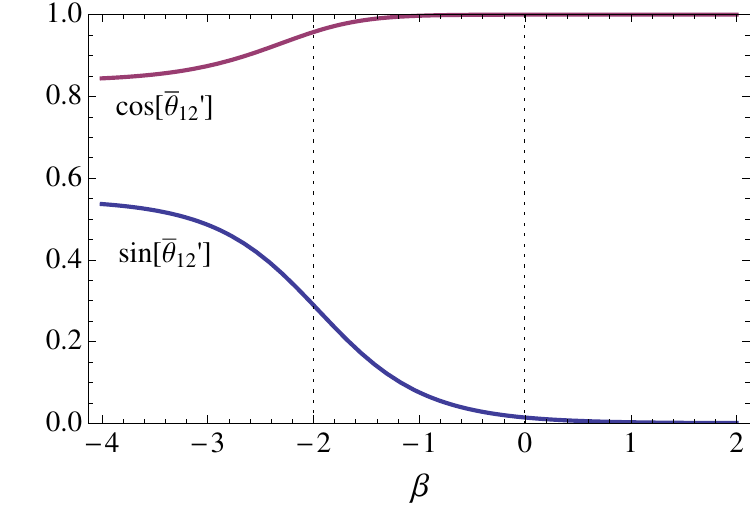}}
\subfigure[]{\includegraphics[height=5cm]{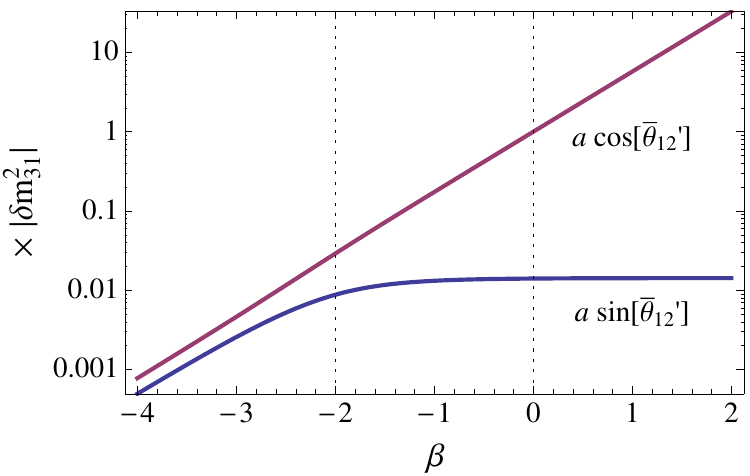}}
\caption{(a) The $\beta$-dependence of $\overline{s}'_{12}=\sin\overline{\theta}'_{12}$ and
$\overline{c}'_{12}=\cos\overline{\theta}'_{12}$.
(b) The $\beta$-dependence of $a\overline{s}'_{12}$ and $a\overline{c}'_{12}$.
The asymptotic value of $a\overline{s}'_{12}$ is $\delta m^2_{21}s_{12}c_{12}/c_{13}^2
\approx 0.014\,|\delta m^2_{31}| = O(\varepsilon^2|\delta m^2_{31}|)$.
}
\label{s12primebarc12primebarplot}
\end{figure}

Define
\begin{equation}
\overline{V} \;\equiv\; 
\left[
\begin{array}{ccc}
\overline{c}_{\varphi} & \overline{s}_{\varphi} & 0 \\
-\overline{s}_{\varphi} & \overline{c}_{\varphi} & 0 \\
0 & 0 & 1
\end{array}
\right] \;,
\end{equation}
where 
\begin{equation}
\overline{c}_{\varphi} \;=\; \cos\overline{\varphi} \;,\quad
\overline{s}_{\varphi} \;=\; \sin\overline{\varphi} \;,\quad
\tan 2\overline{\varphi} \;\equiv\; 
-\dfrac{a c_{13}^2 \sin 2\theta_{12}}
      {\delta m^2_{21} + a c_{13}^2 \cos 2\theta_{12}} \;,
\quad
\left(-\dfrac{\pi}{2}<\overline{\varphi}<0\right)
\;.
\end{equation}
Using $\overline{V}$ we find
\begin{equation}
\overline{H}''_a
\;\equiv\; 
\overline{V}^\dagger \overline{H}'_a \overline{V} 
\;=\; 
\left[
\begin{array}{ccc}
      \overline{\lambda}'_- 
	& 0 
	& -a\overline{c}'_{12} c_{13} s_{13} \\
      0 
	& \overline{\lambda}'_+ 
	& -a \overline{s}'_{12} c_{13} s_{13} \\
      -a \overline{c}'_{12} c_{13} s_{13} 
	& -a \overline{s}'_{12} c_{13} s_{13} 
	& -a s_{13}^2 + \delta m_{31}^2
\end{array}
\right] 
\;,
\label{Hbardoubleprime}
\end{equation}
where
\begin{equation}
\overline{c}'_{12} \;=\; \cos \overline{\theta}'_{12} \;, \quad
\overline{s}'_{12} \;=\; \sin \overline{\theta}'_{12} \;, \quad
\overline{\theta}'_{12} \;\equiv\; \theta_{12} + \overline{\varphi} \;,
\end{equation}
and
\begin{equation}
\overline{\lambda}'_{\pm} 
\;\equiv\; 
\dfrac{ (\delta m_{21}^2 - a c_{13}^2) \pm 
	    \sqrt{ (\delta m_{21}^2 + a c_{13}^2)^2  
		  - 4a c_{13}^2 s_{12}^2 \delta m_{21}^2  }
	  }{2} \;.
\end{equation}
The angle $\overline{\theta}'_{12}$ can be calculated directly without going through
$\overline{\varphi}$ via
\begin{equation}
\tan 2 \overline{\theta}'_{12} 
\;=\;
\dfrac{\delta m_{21}^2 \sin 2\theta_{12} }
	  {\delta m_{21}^2 \cos 2\theta_{12} + a c_{13}^2 } 
\;,\qquad
\left(0\le\overline{\theta}'_{12}\le\theta_{12}\right)
\;.
\end{equation}
The $\beta$-dependences of $\overline{\theta}'_{12}$ and $\overline{\lambda}'_\pm$ are
shown in Fig.~\ref{theta12primebarplot}. 
Note that in contrast to the neutrino case, there is no level crossing.
$\overline{\theta}'_{12}$ decreases monotonically toward zero as $a$ is increased.
For $a\gg\delta m^2_{21}$, 
$\overline{s}'_{12}$ and $\overline{c}'_{12}$ behave as
\begin{eqnarray}
\overline{s}'_{12} 
& = & 
s_{12}c_{12}\left(\dfrac{\delta m^2_{21}}{a c^2_{13}}\right)
- s_{12}c_{12}(c^2_{12}-s^2_{12})\left(\dfrac{\delta m^2_{21}}{a c^2_{13}}\right)^2
+ \cdots
\;,
\cr
\overline{c}'_{12}
& = & 1
- \dfrac{s^2_{12}c^2_{12}}{2}\left(\dfrac{\delta m^2_{21}}{a c^2_{13}}\right)^2
+ \cdots
\;,
\label{sbar12primecbar12prime}
\end{eqnarray}
and we see that, this time, 
we have $a \overline{c}'_{12}\approx a$ and 
$a \overline{s}'_{12}\approx \delta m^2_{21}s_{12}c_{12}/c^2_{13}=O(\varepsilon^2|\delta m^2_{31}|)$.
These $\beta$-dependences of $\overline{s}'_{12}$, $\overline{c}'_{12}$,
$a\overline{s}'_{12}$, and $a\overline{c}'_{12}$ are shown in Fig.~\ref{s12primebarc12primebarplot}(a) and (b).

In the range $a\ll \delta m^2_{21}$, $\overline{\lambda}'_\pm$ can 
be expanded as
\begin{eqnarray}
\overline{\lambda}'_-
& = &
-a c_{13}^2 c_{12}^2
\left[
1
+ s_{12}^2 \left(\dfrac{a c_{13}^2}{\delta m^2_{21}}\right)
- s_{12}^2 (c_{12}^2-s_{12}^2)
\left(\dfrac{a c_{13}^2}{\delta m^2_{21}}\right)^2
+ \cdots
\right]
\;,
\cr
\overline{\lambda}'_+
& = &
\delta m^2_{21}
\left[ 1
- s_{12}^2 \left(\dfrac{a c_{13}^2 }{\delta m^2_{21}}\right)
+ s_{12}^2 c_{12}^2 \left(\dfrac{a c_{13}^2 }{\delta m^2_{21}}\right)^2
- s_{12}^2 c_{12}^2 (c_{12}^2-s_{12}^2)  
\left(\dfrac{a c_{13}^2 }{\delta m^2_{21}}\right)^3
+ \cdots 
\right]
\;,\cr
& &
\end{eqnarray}
while in the range $a\gg \delta m^2_{21}$, we obtain
\begin{eqnarray}
\overline{\lambda}'_-
& = &
-a c_{13}^2
\left[
1
- s_{12}^2 \left(\dfrac{\delta m^2_{21}}{a c_{13}^2}\right)
+ s_{12}^2 c_{12}^2
\left(\dfrac{\delta m^2_{21}}{a c_{13}^2}\right)^2
- s_{12}^2 c_{12}^2
\left(\dfrac{\delta m^2_{21}}{a c_{13}^2}\right)^3
+ \cdots
\right]
\;,
\cr
\overline{\lambda}'_+
& = &
\delta m^2_{21} c_{12}^2
\left[ 
1
+ s_{12}^2 \left(\dfrac{\delta m^2_{21}}{a c_{13}^2 }\right)
- s_{12}^2 (c_{12}^2-s_{12}^2) \left(\dfrac{\delta m^2_{21}}{a c_{13}^2 }\right)^2
+ \cdots 
\right]
\;.
\label{lambdaprimebarexpansion}
\end{eqnarray}
The asymptotic values are thus
$\overline{\lambda}'_+\rightarrow \delta m^2_{21}$,
$\overline{\lambda}'_-\rightarrow -a c^2_{13}c^2_{12}$, 
in the limit $a\rightarrow 0$, and 
$\overline{\lambda}'_+\rightarrow c^2_{12}\delta m^2_{21}$,
$\overline{\lambda}'_-\rightarrow -a c^2_{13}$, 
in the limit $a\rightarrow \infty$.

\subsubsection{Second Rotation}

After the first rotation, the effective hamiltonian was given by Eq.~(\ref{Hbardoubleprime}).
When $a<\delta m^2_{21}$, both 
non-zero off-diagonal elements are of order 
$O(\varepsilon a)<O(\varepsilon^3|\delta m^2_{31}|)$.
In contrast to the neutrino case,
as $a$ increases beyond $\delta m^2_{21}$, 
the angle $\overline{\theta}'_{12}$ approaches $0$, 
and it is the 1-3 element that becomes the larger of the two. 
Therefore, a 1-3 rotation is needed next.

\begin{figure}[t]
\subfigure[]{\includegraphics[height=5.1cm]{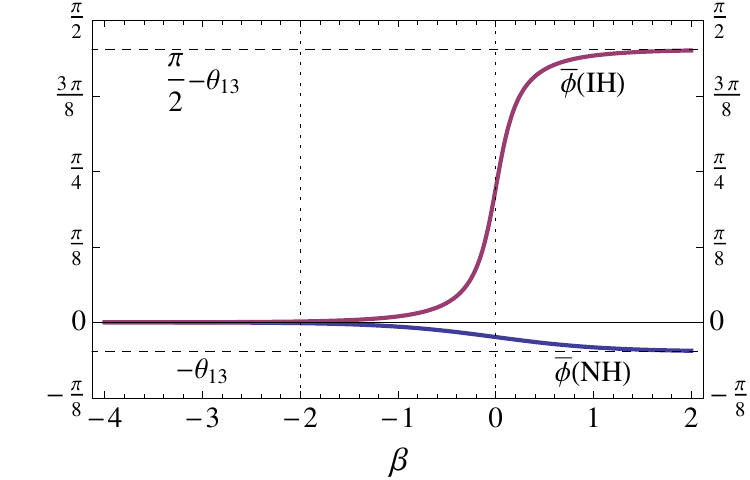}}
\subfigure[]{\includegraphics[height=5cm]{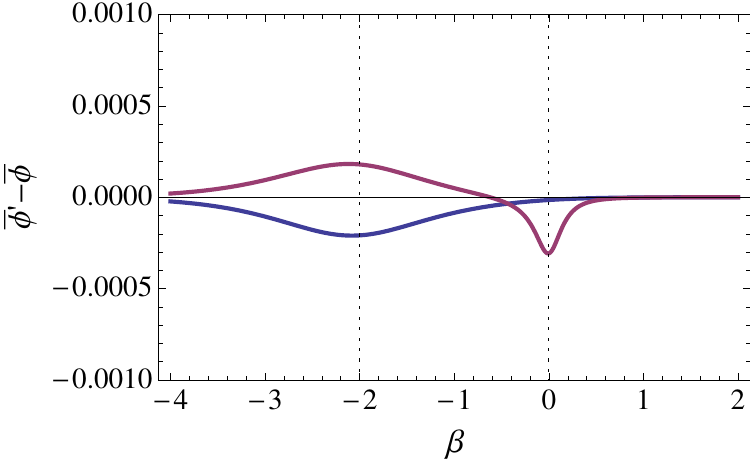}}
\caption{(a) The $\beta$-dependence of $\overline{\phi}$ for the normal and inverted hierarchies.
(b) The $\beta$-dependence of the difference $\overline{\phi}'-\overline{\phi}$.
}
\label{phibarphibarprimecomparison}
\end{figure}

We define
\begin{equation}
\overline{W} 
\;=\;
\left[
	\begin{array}{ccc}
	\overline{c}_{\phi} & 0 & \overline{s}_{\phi} \\
	0 & 1 & 0 \\
	-\overline{s}_{\phi} & 0 & \overline{c}_{\phi}
	\end{array}
\right]
\;,
\end{equation}
where
\begin{equation}
\overline{c}_{\phi} \;=\; \cos \overline{\phi} \;,\quad
\overline{s}_{\phi} \;=\; \cos \overline{\phi} \;,\quad
\tan 2 \overline{\phi} 
\;\equiv\; 
-\dfrac{ a \overline{c}'_{12} \sin 2 \theta_{13} }
	  { \delta m_{31}^2 - a s_{13}^2 - \overline{\lambda}'_- } 
\;.
\label{phibardef}
\end{equation}
The angle $\overline{\phi}$ is in the fourth quadrant when $\delta m_{31}^2 > 0$, 
and the first quadrant when $\delta m_{31}^2 < 0$. 
Using $\overline{W}$, we find
\begin{equation}
\overline{H}'''_a
\;\equiv\; 
\overline{W}^\dagger \overline{H}''_a \overline{W} 
\;=\; 
\left[
\begin{array}{ccc}
	\overline{\lambda}''_\mp 
  & a \overline{s}'_{12} c_{13} s_{13} \overline{s}_{\phi} 
  & 0 
  \\
	a \overline{s}'_{12} c_{13} s_{13} \overline{s}_{\phi} 
  & \overline{\lambda}'_+ 
  & -a \overline{s}'_{12} c_{13} s_{13} \overline{c}_{\phi} 
  \\
	0 
  & -a \overline{s}'_{12} c_{13} s_{13} \overline{c}_{\phi} 
  & \overline{\lambda}''_\pm
\end{array}
\right] \;,
\label{Hbartripleprime}
\end{equation}
where the upper(lower) sign corresponds to normal(inverted) mass hierarchy
with
\begin{equation}
\overline{\lambda}''_{\pm} 
\;\equiv\;
\dfrac{ \bigl[\;\overline{\lambda}'_- + ( \delta m_{31}^2 - a s_{13}^2 ) \,\bigr]
\pm \sqrt{ \bigl[\;\overline{\lambda}'_- - ( \delta m_{31}^2 - a s_{13}^2 ) \,\bigr]^2 
					+ 4 a^2 \overline{c}^{\prime 2}_{12} c_{13}^2 s_{13}^2 } }{2} 
\;.
\end{equation}
The $\beta$-dependence of $\overline{\lambda}''_\pm$ and $\overline{\phi}$ 
are shown in Fig.~\ref{lambdadoubleprimeplot} ((c) and (d)),
and Fig.~\ref{phibarphibarprimecomparison}(a), respectively,
for both normal and inverted mass hierarchies.
For the normal hierarchy case, $\delta m^2_{31}>0$, there 
is no level crossing, and $\overline{\lambda}''_\pm$ are well approximated
by
\begin{equation}
\overline{\lambda}''_+ \;\approx\; \delta m^2_{31}\;,\qquad 
\overline{\lambda}''_- \;\approx\; \overline{\lambda}'_-\;.
\end{equation}
Level crossing occurs for the inverted hierarchy case, $\delta m^2_{31}<0$,
in which we have
\begin{eqnarray}
\overline{\lambda}''_+ & \approx & \overline{\lambda}'_-  \;,\cr
\overline{\lambda}''_- & \approx & -\delta m^2_{31}-a s^2_{13}\;,
\end{eqnarray}
when $a\ll \delta m^2_{31}$, and
\begin{eqnarray}
\overline{\lambda}''_+ & \approx & -c^2_{13}\delta m^2_{31}+s^2_{13}s^2_{12}\delta m^2_{21}\;,\cr
\overline{\lambda}''_- & \approx & -a -s^2_{13}\delta m^2_{31} +c^2_{13}s^2_{12}\delta m^2_{21}\;,
\end{eqnarray}
when $a\gg \delta m^2_{31}$.

Here, as in the neutrino case, we approximate $\overline{\phi}$
with the angle $\overline{\phi}'$ defined via
\begin{equation}
\tan 2\overline{\phi}'
\;=\; -\dfrac{a\sin 2\theta_{13}}{(\delta m^2_{31}-s^2_{12}\delta m^2_{21})+a\cos^2 2\theta_{13}}
\;,
\end{equation}
which is obtained by using Eqs.~(\ref{sbar12primecbar12prime}) and (\ref{lambdaprimebarexpansion}) on Eq.~(\ref{phibardef}).
The difference between $\overline{\phi}'$ and $\overline{\phi}$ is
shown in Fig.~\ref{phibarphibarprimecomparison}(b), and it is clear that
the difference is negligible.

%

Now, the effective Hamiltonian after the second rotation was 
given by Eq.~(\ref{Hbartripleprime}).
Note that all of the non-zero off-diagonal elements include the factor $a \overline{s}'_{12}$,
which is never larger than $O(\varepsilon^2|\delta m^2_{31}|)$ regardless of the value of $a$ as discussed above.
They also all include a factor of $s_{13}$, which is $O(\varepsilon)$ as we have seen in Eq.~(\ref{scorders}).
Therefore, all off-diagonal elements of $\overline{H}_a'''$ are of order 
$O(\varepsilon^2 s_{13}|\delta m^2_{31}|)=O(\varepsilon^3|\delta m^2_{31}|)$
or smaller regardless of the size of $a$.
We conclude that, at this point, 
off-diagonal elements are negligible and a third rotation is not necessary.

\subsubsection{Absorption of $\overline{\phi}'$ into $\theta_{13}$}

From the above consideration, we conclude that the matrix which diagonalizes $\overline{H}_a'$, Eq.~(\ref{Hbarprime}), is given approximately by $\overline{V}\overline{W}$, and 
that the effective anti-neutrino mixing matrix becomes
\begin{equation}
\edlit{U}
\;\approx\;  U\mathcal{Q}\overline{V}\overline{W} 
\;=\;
\underbrace{R_{23}(\theta_{23},0)R_{13}(\theta_{13},\delta)R_{12}(\theta_{12},0)}_{\displaystyle U}
\mathcal{Q}
\underbrace{R_{12}(\overline{\varphi,0})}_{\displaystyle \overline{V}}
\underbrace{R_{13}(\overline{\phi}',0)}_{\displaystyle \overline{W}} 
\;.
\end{equation}
As in the neutrino case,
we find
\begin{eqnarray}
\edlit{U}
& \approx & 
R_{23}(\theta_{23},0)R_{13}(\theta_{13},\delta)R_{12}(\theta_{12},0)
\mathcal{Q}\,R_{12}(\overline{\varphi},0)R_{13}(\overline{\phi}',0) 
\cr
& = & 
R_{23}(\theta_{23},0)\mathcal{Q}\,
R_{13}(\theta_{13},0)R_{12}(\theta_{12},0)R_{12}(\overline{\varphi},0)R_{13}(\overline{\phi}',0) 
\cr
& = & 
R_{23}(\theta_{23},0)\mathcal{Q}\,
R_{13}(\theta_{13},0)R_{12}(\theta_{12}+\overline{\varphi},0)R_{13}(\overline{\phi}',0) 
\cr
& = & 
R_{23}(\theta_{23},0)\mathcal{Q}\,
R_{13}(\theta_{13},0)R_{12}(\overline{\theta}'_{12},0)R_{13}(\overline{\phi}',0) 
\;.
\end{eqnarray}
Here, we argue that
\begin{equation}
R_{12}(\overline{\theta}'_{12},0)R_{13}(\overline{\phi}',0) 
\;\approx\;
R_{13}(\overline{\phi}',0)R_{12}(\overline{\theta}'_{12},0) 
\;,
\label{RcommuteAnti}
\end{equation}
that is, the 1-3 rotation passes through $R_{12}(\overline{\theta}'_{12},0)$.
This is due to the fact that $\overline{\phi}'$ only becomes non-negligible
when $a\gg \delta m^2_{12}$ where $\overline{s}'_{12}\approx 0$ and $\overline{c}'_{12}\approx 1$,
which means
\begin{equation}
R_{12}(\overline{\theta}'_{12},0)\;\approx\;
\left[\begin{array}{ccc}
 1 & 0 & 0 \\
 0 & 1 & 0 \\  
 0 & 0 & 1 
\end{array}\right]
\;,
\end{equation}
thus any matrix will commute with $R_{12}(\overline{\theta}'_{12},0)$.
In the range $a\alt \delta m^2_{21}$, the angle
$\phi'$ is very small and both $R_{23}(\phi',0)$
and $R_{13}(\phi',0)$ are approximately unit matrices and Eq.~(\ref{RcommuteAnti})
is trivially satisfied.
The accuracy of this approximation is discussed in appendix~\ref{sec:Commute}.
Therefore,
\begin{eqnarray}
\tilde{U}
& \approx & 
R_{23}(\theta_{23},0)\mathcal{Q}\,
R_{13}(\theta_{13},0)R_{12}(\overline{\theta}'_{12},0)R_{13}(\overline{\phi}',0) 
\cr
& \approx & 
R_{23}(\theta_{23},0)\mathcal{Q}\,
R_{13}(\theta_{13},0)R_{13}(\overline{\phi}',0)R_{12}(\overline{\theta}'_{12},0) 
\cr
& = & 
R_{23}(\theta_{23},0)\mathcal{Q}\,
R_{13}(\theta_{13}+\overline{\phi}',0)R_{12}(\overline{\theta}'_{12},0) 
\cr
& = & 
R_{23}(\theta_{23},0)\mathcal{Q}\,
R_{13}(\overline{\theta}'_{13},0)R_{12}(\overline{\theta}'_{12},0) 
\cr
& = & 
R_{23}(\theta_{23},0)
R_{13}(\overline{\theta}'_{13},\delta)R_{12}(\overline{\theta}'_{12},0) 
\mathcal{Q}
\;,
\end{eqnarray}
where we have defined
\begin{equation}
\overline{\theta}'_{13} \;\equiv\; \theta_{13}+\overline{\phi}'\;.
\end{equation}
This angle can be calculated directly without calculation $\overline{\phi}'$ via
\begin{equation}
\tan 2\overline{\theta}'_{13} \;=\;
\dfrac{(\delta m^2_{31}-\delta m^2_{21} s_{12}^2)\sin 2\theta_{13}}
{(\delta m^2_{31}-\delta m^2_{21} s_{12}^2)\cos 2\theta_{13}+a}
\;.
\end{equation}
The phase matrix $\mathcal{Q}$ appearing rightmost in the above matrix product
can be absorbed into the redefinition of the major phases and can be dropped.
Thus, we arrive at our final approximation in which the vacuum mixing angles
are replaced by their effective values in matter
\begin{eqnarray}
\theta_{12} & \rightarrow & \overline{\theta}'_{12} \;=\; \theta_{12}+\overline{\varphi}\;,\cr
\theta_{13} & \rightarrow & \overline{\theta}'_{13} \;=\; \theta_{13}+\overline{\phi}'\;,\cr
\theta_{23} & \rightarrow & \theta_{23}\;,\cr
\delta      & \rightarrow & \delta\;,
\end{eqnarray}
and the eigenvalues of the effective Hamiltonian are given by 
\begin{eqnarray}
\overline{\lambda}_{1} & \approx & \overline{\lambda}''_{\mp}\;,\cr
\overline{\lambda}_{2} & \approx & \overline{\lambda}'_{+}\;,\cr
\overline{\lambda}_{3} & \approx & \overline{\lambda}''_{\pm}\;.
\end{eqnarray}
Note that of the mixing angles, only $\theta_{12}$ and $\theta_{13}$ are shifted.
$\theta_{23}$ and $\delta$ stay at their vacuum values.

\section{Commutation of $R_{13}$ and $R_{23}$ through $R_{12}$}
\label{sec:Commute}

In the derivation of our approximation formulae above, Eqs.~(\ref{Rcommute}) and
(\ref{RcommuteAnti}) played crucial roles in allowing the second rotation angle
to be absorbed into $\theta_{13}$.  In this appendix, we evaluate the validity of
these approximations.

\subsection{Neutrino Case}

\begin{figure}[t]
\subfigure[Normal Hierarchy]{\includegraphics[height=4.5cm]{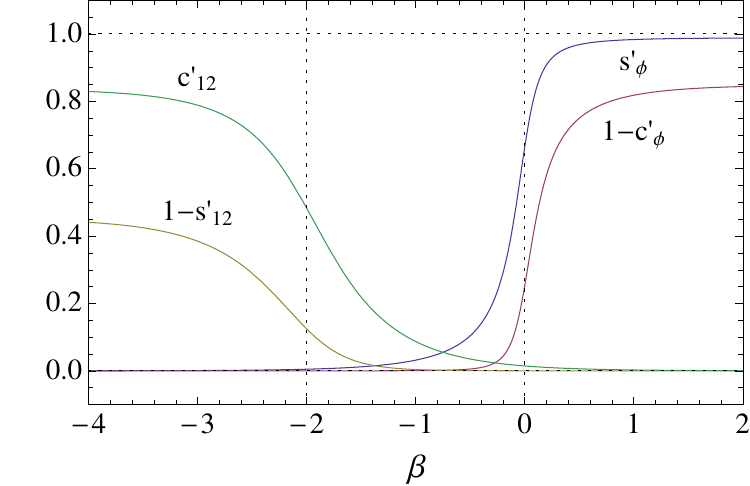}}
\subfigure[Inverted Hierarchy]{\includegraphics[height=4.5cm]{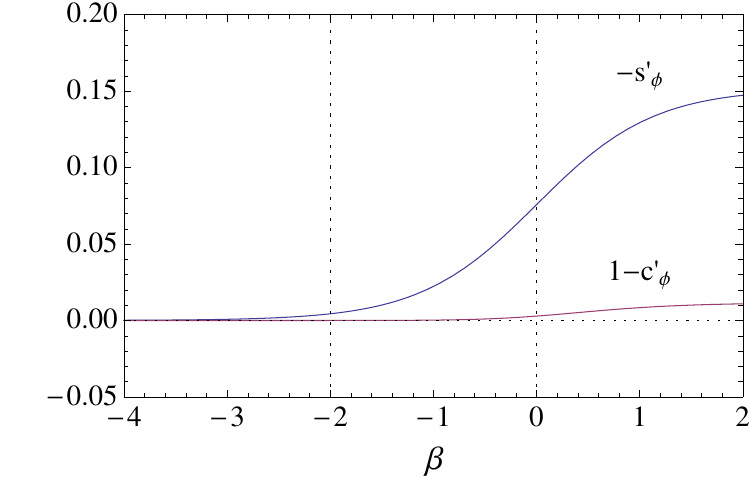}}
\caption{$\beta$-dependence of
$c'_{12}$, $1-s'_{12}$, $s'_\phi$ and $1-c'_{\phi}$ for
(a) normal and (b) inverted hierarchies.
The behaviors of $c'_{12}$ and $1-s'_{12}$ are common
to both.
}
\label{fig:Rcommute1}
\end{figure}

\begin{figure}[t]
\subfigure[Normal Hierarchy]{\includegraphics[height=4.5cm]{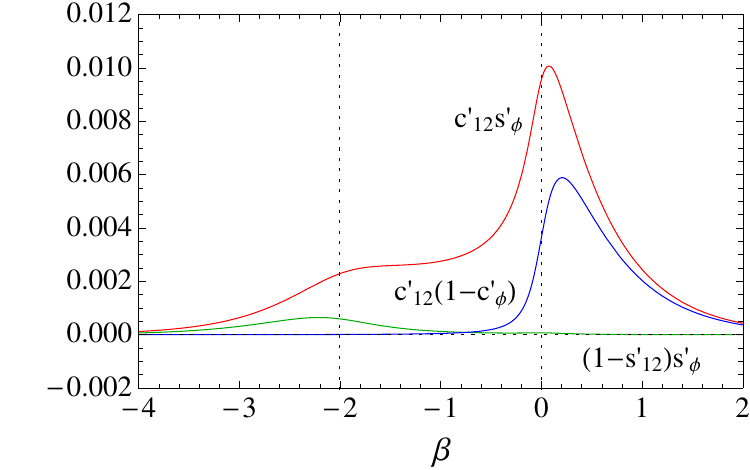}}
\subfigure[Inverted Hierarchy]{\includegraphics[height=4.5cm]{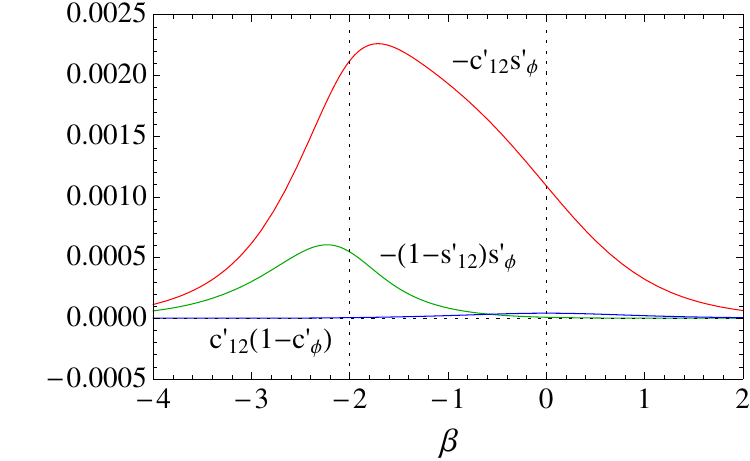}}
\caption{$\beta$-dependence of
the non-zero elements of $\delta R$ for 
(a) normal and (b) inverted hierarchies.
}
\label{fig:Rcommute2}
\end{figure}

The difference between the two sides of Eq.~(\ref{Rcommute}) is given by
\begin{eqnarray}
\delta R
& \equiv & R_{12}( \theta_{12}' , 0 ) R_{23}( \phi', 0 ) - R_{13}( \phi' , 0 ) R_{12}( \theta_{12}' , 0 ) 
\phantom{\bigg|}
\cr
& = &
\left[								
	\begin{array}{ccc}
	c_{12}' (1-c'_{\phi}) & 0                       & -(1 - s_{12}')s'_{\phi} \\
	0                    & -c_{12}' (1 - c'_{\phi}) & c_{12}' s'_{\phi}       \\
    c_{12}' s'_{\phi}     & -(1 - s_{12}')s'_{\phi}  & 0
	\end{array}
\right] 
\;.
\end{eqnarray}
It is clear that $\delta R$ will vanish 
in the two limits $a\rightarrow 0$ where 
$s'_{12}\rightarrow s_{12}$, $c'_{12}\rightarrow c_{12}$,
$s'_{\phi}\rightarrow 0$, and $c'_{\phi}\rightarrow 1$,
and $a\rightarrow\infty$ where
$s'_{12}\rightarrow 1$, $c'_{12}\rightarrow 0$, 
$s'_{\phi}\rightarrow c_{13}(-s_{13})$, and $c'_{\phi}\rightarrow s_{13}(c_{13})$
for normal(inverted) hierarchy.
The question is whether $\delta R$ will stay negligible in between as
$s'_{12}$ runs from $s_{12}$ to $1$, $c'_{12}$ from $c_{12}$ to $0$,
$s'_{\phi}$ from $0$ to $c_{13}$ (normal) or $-s_{13}$ (inverted), and
$c'_{\phi}$ from $1$ to $s_{13}$ (normal) or $c_{13}$ (inverted)
as shown in Fig.~\ref{fig:Rcommute1}.
The dependence of the non-zero elements of $\delta R$
on $\beta=-\log_\varepsilon(a/|\delta m^2_{31}|)$ 
is shown in Fig.~\ref{fig:Rcommute2}.
The bumps at $a\sim\delta m^2_{21}$ for both hierarchies, and
that at $a\sim\delta m^2_{31}$ for the normal hierarchy, happen
due to the $\theta'_{12}$ factor competing with the $\phi'$ factor
as one of them goes through a resonance while the other damps to zero.
The heights of the bumps depend on the narrowness of the resonances.

\begin{figure}[t]
\subfigure[$s_{13}=0.03$, Normal Hierarchy]{\includegraphics[height=5cm]{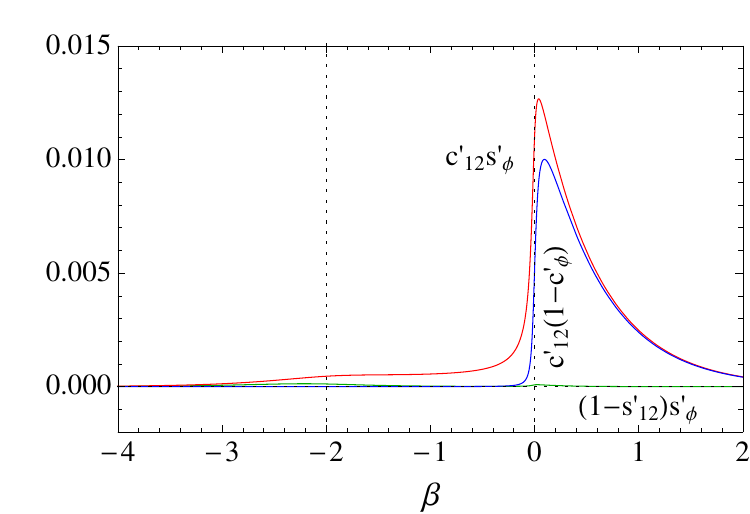}}
\subfigure[$s_{13}=0.005$, Normal Hierarchy]{\includegraphics[height=5cm]{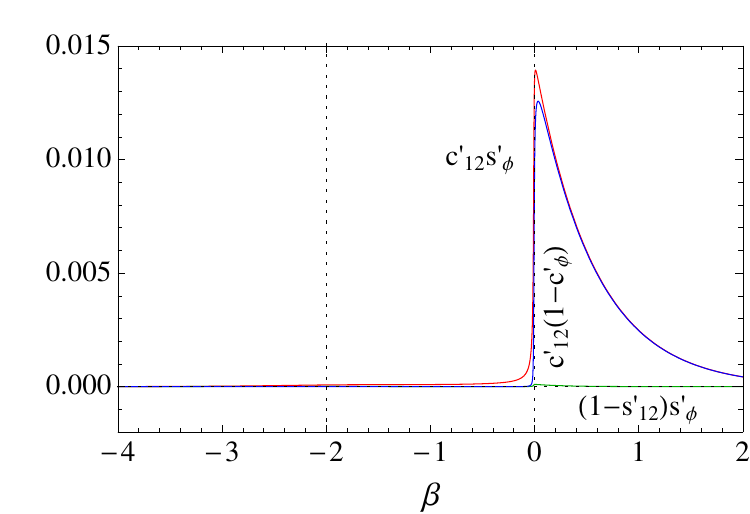}}
\caption{$\beta$-dependence of
the non-zero elements of $\delta R$ for 
different values of $s_{13}$ with normal hierarchy.
(a) $s_{13}=0.03=O(\varepsilon^2)$, (b) $s_{13}=0.005=O(\varepsilon^3)$.
}
\label{fig:Rcommute3}
\end{figure}

For the case shown in Fig.~\ref{fig:Rcommute2},
which was generated with the numbers in Table~\ref{tab:bench} as input,
all elements of $\delta R$ are $O(\varepsilon^3)$ or smaller for the
entire range of $a$, 
with the maximum value of $\sim 0.01 \approx 2\varepsilon^3$
occurring in $c'_{12}s'_{\phi}$ near $a\sim\delta m^2_{31}$ in the normal hierarchy case.
Since the size of the third rotation angle we neglected in the
Jacobi procedure was $O(\varepsilon^2 s_{13})$, 
Eq.~(\ref{Rcommute}) is valid to the same order provided $s_{13}=O(\varepsilon)$.

For smaller values of $s_{13}$, the resonance at $a\sim\delta m^2_{31}$ would
have been narrower, and the peaks in $c'_{12}s'_{\phi}$ and $c'_{12}(1-c'_{\phi})$ higher.
This is illustrated in Fig.~\ref{fig:Rcommute3}.
In the limit $s_{13}\rightarrow +0$, $s'_{\phi}$ and $1-c'_{\phi}$ will
become step functions at $\beta\sim 0$, and the maximum height of the peak will
be
\begin{equation}
c'_{12}(a\sim\delta m^2_{31}) \;\approx\; s_{12}c_{12}\varepsilon^2
\;=\; 0.46\,\varepsilon^2
\;=\; 0.014 
\;,
\label{peakheight}
\end{equation}
as can be discerned from Eq.~(\ref{s12primec12prime}).
This is the same as the asymptotic value of $a c'_{12}/\delta m^2_{31}$ discussed earlier.
While this value may not seem particularly large, only a factor of $3/2$ larger than the peak
in Fig.~\ref{fig:Rcommute2}(a),
it is parametrically $O(\varepsilon^2)$.
On the other hand, the third rotation angle neglected in the Jacobi procedure was $O(\varepsilon^2 s_{13})$.
Thus, using Eq.~(\ref{Rcommute}) would lead to dropping terms that are larger than the
ones we keep when $s_{13}=O(\varepsilon^2)$ or smaller.
Also, the sudden change in the accuracy of Eq.~(\ref{Rcommute}) across
$a\sim\delta m^2_{31}$, as can be seen
in Fig.~\ref{fig:Rcommute3}, will lead to kinks in the resulting oscillation probabilities.

\subsection{Anti-neutrino Case}

\begin{figure}[t]
\subfigure[Normal Hierarchy]{\includegraphics[height=4.5cm]{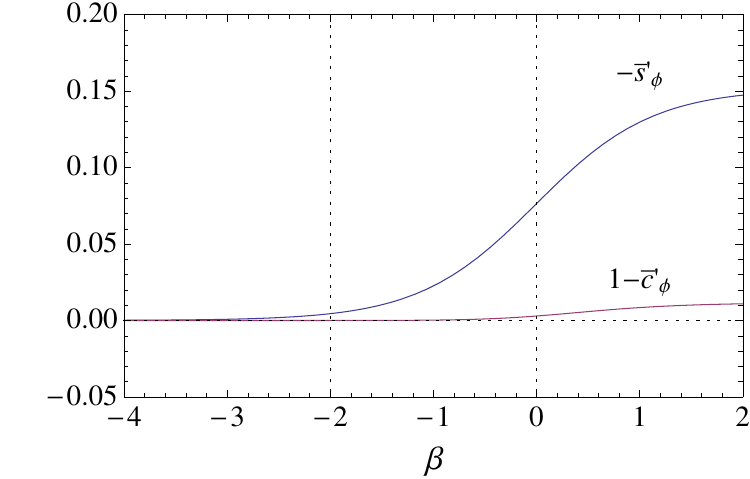}}
\subfigure[Inverted Hierarchy]{\includegraphics[height=4.5cm]{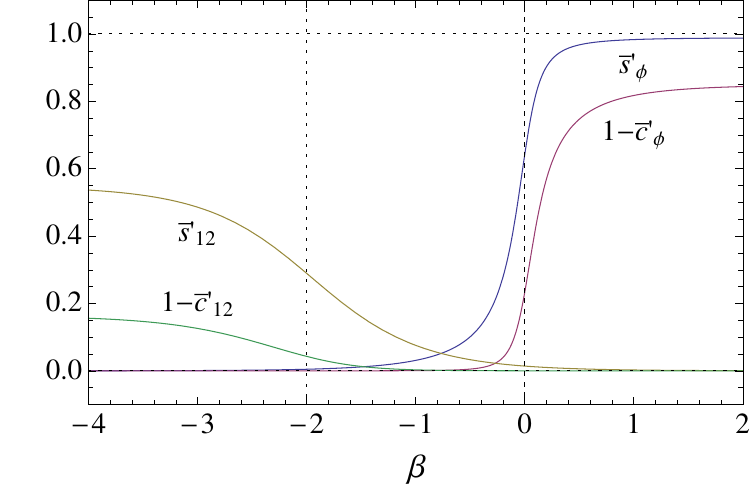}}
\caption{$\beta$-dependence of
$\overline{s}'_{12}$, $1-\overline{c}'_{12}$, $s'_\phi$ and $1-c'_{\phi}$ for
(a) normal and (b) inverted hierarchies.
The behaviors of $\overline{s}'_{12}$ and $1-\overline{c}'_{12}$ are common
to both.
}
\label{fig:Rcommute1bar}
\end{figure}

The difference between the two sides of Eq.~(\ref{RcommuteAnti}) is given by
\begin{eqnarray}
\overline{\delta R}
& \equiv & R_{12}(\overline{\theta}_{12}' , 0 ) R_{13}(\overline{\phi}', 0 ) 
- R_{13}( \overline{\phi}' , 0 ) R_{12}( \overline{\theta}_{12}' , 0 ) 
\phantom{\bigg|}
\cr
& = &
\left[								
	\begin{array}{ccc}
	0 & \overline{s}_{12}' (1-\overline{c}'_{\phi}) & -(1 - \overline{c}_{12}')\overline{s}'_{\phi} \\
	\overline{s}_{12}' (1-\overline{c}'_{\phi}) & 0 & -\overline{s}_{12}' \overline{s}'_{\phi} \\
    (1 - \overline{c}_{12}')\overline{s}'_{\phi} & \overline{s}_{12}' \overline{s}'_{\phi} & 0
	\end{array}
\right] 
\;.
\end{eqnarray}
It is clear that $\overline{\delta R}$ will vanish 
in the two limits $a\rightarrow 0$ where 
$\overline{s}'_{12}\rightarrow s_{12}$, $\overline{c}'_{12}\rightarrow c_{12}$,
$\overline{s}'_{\phi}\rightarrow 0$, and $\overline{c}'_{\phi}\rightarrow 1$,
and $a\rightarrow\infty$ where
$\overline{s}'_{12}\rightarrow 0$, $\overline{c}'_{12}\rightarrow 1$, 
$\overline{s}'_{\phi}\rightarrow -s_{13}(c_{13})$, and 
$\overline{c}'_{\phi}\rightarrow c_{13}(s_{13})$
for normal(inverted) hierarchy.
The question is whether $\delta R$ will stay negligible in between as
$\overline{s}'_{12}$ runs from $s_{12}$ to $0$, $\overline{c}'_{12}$ from $c_{12}$ to $1$,
$\overline{s}'_{\phi}$ from $0$ to $-s_{13}$ (normal) or $c_{13}$ (inverted), and
$\overline{c}'_{\phi}$ from $1$ to $c_{13}$ (normal) or $s_{13}$ (inverted)
as shown in Fig.~\ref{fig:Rcommute1bar}.

\begin{figure}[t]
\subfigure[Normal Hierarchy]{\includegraphics[height=4.5cm]{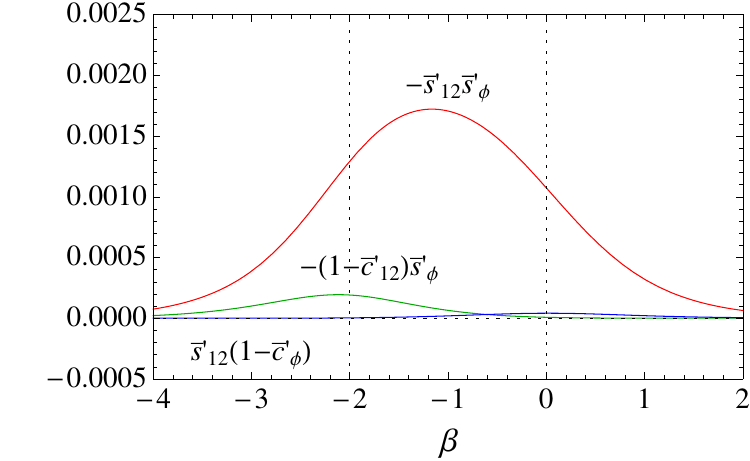}}
\subfigure[Inverted Hierarchy]{\includegraphics[height=4.5cm]{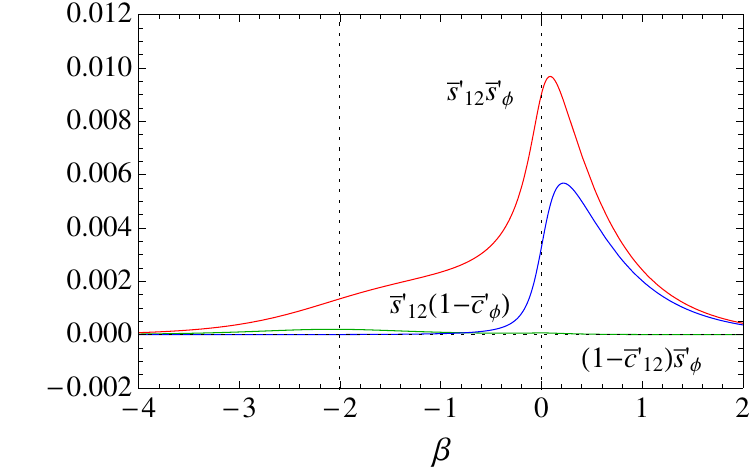}}
\caption{$\beta$-dependence of
the non-zero elements of $\overline{\delta R}$ for 
(a) normal and (b) inverted hierarchies.
}
\label{fig:Rcommute2bar}
\end{figure}
\begin{figure}[t]
\subfigure[$s_{13}=0.03$, Inverted Hierarchy]{\includegraphics[height=5cm]{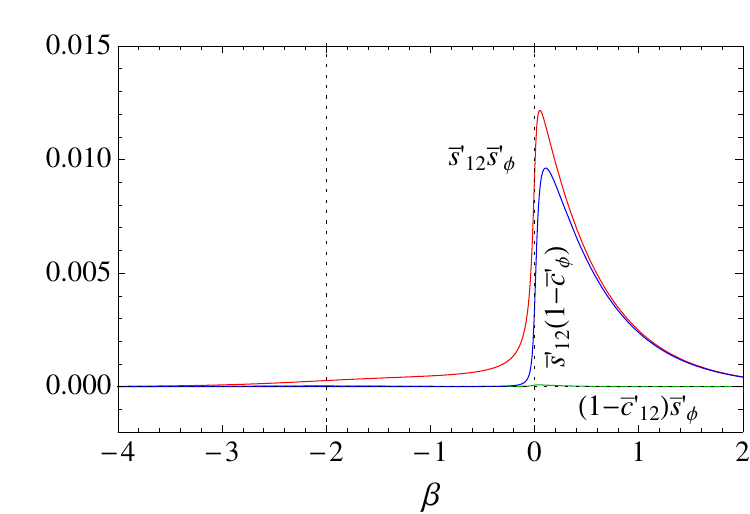}}
\subfigure[$s_{13}=0.005$, Inverted Hierarchy]{\includegraphics[height=5cm]{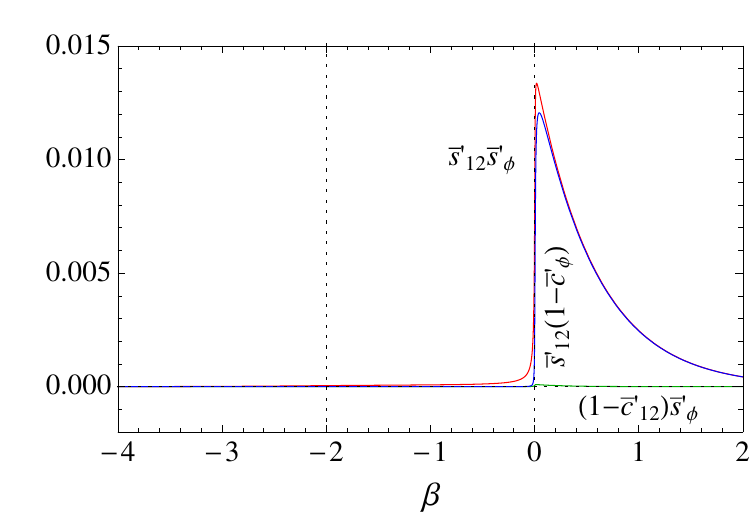}}
\caption{$\beta$-dependence of
the non-zero elements of $\overline{\delta R}$ for 
different values of $s_{13}$ with inverted hierarchy.
(a) $s_{13}=0.03=O(\varepsilon^2)$, (b) $s_{13}=0.005=O(\varepsilon^3)$.
}
\label{fig:Rcommute3bar}
\end{figure}

The dependence of the non-zero elements of $\overline{\delta R}$
on $\beta=-\log_\varepsilon(a/|\delta m^2_{31}|)$ 
is shown in Fig.~\ref{fig:Rcommute2bar},
which was generated with the numbers in Table~\ref{tab:bench} as input.
We can see that all elements of $\overline{\delta R}$ are $O(\varepsilon^3)$ or smaller for the
entire range of $a$, 
with the maximum value of $\sim 0.01 \approx 2\varepsilon^3$
occuring in $\overline{s}'_{12}\overline{s}'_{\phi}$ near $a\sim\delta m^2_{31}$ in the inverted hierarchy case.
Since the size of the third rotation angle we neglected in the
Jacobi procedure was $O(\varepsilon^2 s_{13})$, 
Eq.~(\ref{RcommuteAnti}) is valid to the same order provided $s_{13}=O(\varepsilon)$.

For smaller values of $s_{13}$, the resonance at $a\sim\delta m^2_{31}$ would
have been narrower, and the peaks in $\overline{s}'_{12}\overline{s}'_{\phi}$ and 
$\overline{s}'_{12}(1-\overline{c}'_{\phi})$ higher.
This is illustrated in Fig.~\ref{fig:Rcommute3bar}.
In the limit $s_{13}\rightarrow +0$, $\overline{s}'_{\phi}$ and $1-\overline{c}'_{\phi}$ will
become step functions at $\beta\sim 0$, and the maximum height of the peak will
be the same as Eq.~(\ref{peakheight}), and the asymptotic value of $a\overline{s}'_{12}/|\delta m^2_{31}|$,
as can be discerned from Eq.~(\ref{sbar12primecbar12prime}).
This is parametrically $O(\varepsilon^2)$, while 
the third rotation angle neglected in the Jacobi procedure was $O(\varepsilon^2 s_{13})$.
Thus, using Eq.~(\ref{RcommuteAnti}) would lead to dropping terms that are larger than the
ones we keep when $s_{13}=O(\varepsilon^2)$ or smaller. 
Also, the sudden change in the accuracy of Eq.~(\ref{RcommuteAnti}) across
$a\sim\delta m^2_{31}$, as can be seen
in Fig.~\ref{fig:Rcommute3bar}, will lead to kinks in the resulting oscillation probabilities.

\bibliographystyle{JHEP}
\bibliography{osc-references}

\end{document}